# Efficient Electrocatalytic H$_2$ Evolution Mediated by 2D Janus MoSSe Transition Metal Dichalcogenide

## Srimanta Pakhira[1, 2,3*] and Shrish Nath Upadhyay[2]

[1] Department of Physics, Indian Institute of Technology Indore (IIT Indore), Simrol, Khandwa Road, Indore-453552, Madhya Pradesh, India.
[2] Department of Metallurgy Engineering and Materials Science (MEMS), Indian Institute of Technology Indore, Simrol, Khandwa Road, Indore-453552, Madhya Pradesh, India.
[3] Centre for Advanced Electronics (CAE), Indian Institute of Technology Indore, Simrol, Khandwa Road, Indore-453552, Madhya Pradesh, India.

*Corresponding author: spakhira@iiti.ac.in (or) spakhirafsu@gmail.com

## Abstract

Recently, 2D Janus Transition Metal Dichalcogenides (JTMDs) with asymmetric electronic structures are inviting an intense research interest in modern science and technology. Using the first principles-based periodic hybrid dispersion-corrected Density Functional Theory (DFT-D) method, we have investigated the equilibrium structure, geometry, and electronic properties of the 2D monolayer MoSSe JTMD with the electrocatalytic activities for the H$_2$ evolution reaction (HER). We have performed non-periodic quantum mechanical DFT computations to find out the most favorable HER pathway on the exposed surfaces of the 2D Janus MoSSe material i.e., on the Mo-edges (10$\bar{1}$0) and S-/Se-edges ($\bar{1}$010). To explore the electrocatalytic HER mechanism, reaction pathways and barriers, we have considered a cluster model system Mo$_{10}$S$_{12}$Se$_9$ to illustrate the Mo-edges (10$\bar{1}$0) and S-/Se-edges ($\bar{1}$010) of the 2D monolayer MoSSe material. The present study reveals that the Volmer−Heyrovsky reaction mechanism is thermodynamically favorable reaction pathway to evolute H$_2$ at the Se-terminated Mo-edges. It was found that the change of free energy barrier (ΔG) during the Heyrovsky reaction at the Se-terminated Mo-edges is about 3.93 - 7.10 kcal.mol$^{-1}$ (in both the gas and the solvent phases), indicating an exceptional electrocatalyst for HER with the lowest activation barriers. This study showed that the Tafel slope (*m*) is lower in the case of 2D Janus MoSSe material due to the overlap of the *s*-orbital of the hydrogen and *d*-orbitals of the Mo atoms appeared in the HOMO and LUMO transition state TS1 of the H•-migration reaction step. A better overlap of the *s*-orbitals of the hydrogen atom attached with the Mo and the water cluster ($H_3O^+$ + $3H_2O$) seemed in the HOMO-LUMO Heyrovsky's transition state TS2 has found, and this better overlap of the atomic orbitals during the $H_2$ formation in the Heyrovsky's TS2 reduces the reaction barrier. The better stabilization of the atomic orbitals in the HER rate-limiting step i.e., H•-migration TS1 reaction step (in the solvent phase) is a key for reducing the reaction barrier, thus the overall catalysis indicating a better electrocatalytic performance for $H_2$ evolution. The present investigation implies how to computationally invent a highly reactive 2D electrocatalyst from the JTMDs with their exposed edges which are active for effective HER. This research will enhance the development of 2D JTMDs which will show the lower reaction barriers during $H_2$ evolution.

## Introduction:

The usage of fossil fuels and many non-renewable energy resources are diminishing the environment due to carbon emissions in the atmosphere. To control the pollution in the atmosphere, alternate energy storage devices like alkali-ion batteries, super capacitors, metal-air batteries, porous materials such as PCPs, COFs, MOFs, etc. have shown excellent performance due to their high power density, energy density, etc.[1–7] It is important to find out alternative pollution free energy resources. Hydrogen is a desirable renewable energy resource to replace existing fossil fuels since it contains high energy density in its molecular form.[8] In general, $H_2$ is generated by using steam reforming, partial oxidation, autothermal reforming with the help of hydrocarbon-based fuels, but it also produces carbon emissions as a byproduct, which hinders the concept of clean and renewable energy technology.[9] So, to avoid the emissions of greenhouse gases such as CO, $CO_2$, etc. there are other sustainable technologies to generate hydrogen via electrolysis, thermolysis, and photoelectrolysis.[10,11]

Recently, it has been studied that the electrolysis of water is one of the useful methods to produce $H_2$ with the reasonable production cost, and it is non-pollution process, cost effective, highly efficient and environment friendly.[12,13] Modern research has been focused on the development of earth-abundant, cost-effective, and highly efficient electrocatalysts for the effective $H_2$ evolution reaction (HER: $2H^+ + 2e^- = H_2$). The main key to explore the efficiency of $H_2O$ splitting is the electrocatalytic cathode materials which are required in the electrolysis of $H_2O$. There are a few noble metals like palladium (Pd), platinum (Pt), ruthenium (Ru), etc., shown excellent electrocatalytic activity for HER mechanism.[14,15] Till to date, it is considered that the Pt-based cathode materials are the best electrocatalysts for effective HER due to near-zero overpotential. Due to its high cost and scarce resource, it hinders the practical applications in industries which limit their commercial applications. It is a challenging task to find out an excellent cost-effective and earth-abundant electrocatalyst to obtain hydrogen from the electrolysis of water at a faster rate.[16,17] Therefore, it is an inspirational mission to develop an alternative Pt-free highly efficient electrocatalyst for effective HER which can be applicable for $H_2$-based energy technology. For the development of such kind of electrocatalyst, four conditions must be considered: stability should be high, cost should be low with good electrical conductivity and large specific surface area and multiple active sites, and remarkable activity without creating defects. In this direction, 2D materials could be the most promising candidates for the electrocatalyst as they have large surface areas, which provides more active sites for the reactions, and they are of low cost and have environmentally friendly nature.[18]

Very recently, 2D transition metal dichalcogenides (TMDs) has attracted tremendously in materials chemistry and they have been considerably explored as a potential electrocatalyst for effective $H_2$ evolution due to its exceptional materials structures, hexagonal symmetry, and unique electronic properties.[12] Among all the TMDs, the pure 2D monolayer $MoS_2$ has emerged as an auspicious material for effective $H_2$ evolution in acidic conditions and the electrocatalytic performance of this material is very close to the Pt-based materials.[19] It has been computationally reported that the 2D single layer $MoS_2$ TMD can catalyze electrochemical $H_2$ evolution at a moderate overpotential of 0.1-0.2 V due to its approximately thermoneutral $H_2$ adsorption energy.[20] Huang et al. computationally explored the $H_2$ evolution reaction pathways, mechanism and reaction barriers for the HER on the surfaces of 2D monolayer $MoS_2$ materials by employing first principles-based Density Functional Theory (DFT) method.[21] Recently, Lei et al. synthesized hybrid $W_xMo_{1-x}S_2$/rGO (rGO = reduced graphene oxide) heterostructure films by using a simple wet chemical approach.[12] They found that these TMD alloy (i.e., $W_xMo_{1-x}S_2$) materials are effective electrocatalysts for H$_2$ evolution. Among all the TMD alloys, the heterostructure made by reduced graphene oxide (rGO) and W$_{0.4}$Mo$_{0.6}$S$_2$ alloy material has shown better electrocatalytic efficiency for HER than the pristine 2D monolayer WS$_2$ and MoS$_2$ materials.[12] They carried out a computational investigation by applying M06-L DFT method to analyze their experimental observation by investigating the detailed HER mechanism.

Due to several constraints of TMDs, their overall electrocatalytic performance is limited and the pristine TMDs have inert basal planes and it contains a smaller number of active edge sites which are the main drawbacks for excellent electrocatalytic activity for HER. The key challenge is the phase engineering technique to activate the inert basal planes of the TMDs by introducing more number and electrocatalytic activity of basal plane sites.[22] To overcome the limitations of the pristine TMDs for excellent HER, a new technology is required which can potentially enhance the electrochemical performance of H$_2$ evolution reactions by tuning the electrocatalytic properties of the pristine one. Recently, 2D Janus TMDs have invited extensive interest for the further advancement of nanomaterials science, electrolysis of water, fuel cell and renewable energy technology. In this context, Janus TMDs have shown better electrocatalytic performance for HER mechanism with less application of strain and chalcogen anion vacancy in the structure.[23] It has been studied that the Janus TMDs layers (MXY, where M = transition metal atoms, W; X/Y= VI A group elements) have shown intensification of inactive sites at basal planes as well as edge in contrast to the pristine ordinary TMD (like MX$_2$) layers which helps to increase the evolution of hydrogen.[24,25] Due to the different electronegativity of the X and Y layer of chalcogen atoms in the Janus MXY structure, an inherent dipole moment exists. This property distributes holes and electrons present in its structure resourcefully on its surface for upgrading electrocatalytic activity.[26] It should be mentioned here that the 2D Janus TMD material exhibits an inherent out-of-plane electric field owing to the mirror asymmetry in the z-axis and thus shows large Rashba band splitting and piezoelectricity. The single-layer structures of the pristine 2D Janus TMDs are a direct electronic band gap semiconductor and they are thermodynamically stable.[27,28] A recent computational study reported that the 2D monolayer In$_2$SSe Janus TMD is a pure direct band gap semiconductor and it has hexagonal geometry with *P$_3$m$_1$* symmetry such as 2D monolayer MoS$_2$, WS$_2$, InS, MoSe$_2$ and InSe single layers, and they are all energetically and thermodynamically stable materials.[28] It was found that the polarized 2D Janus MoSSe have exceptional novel properties such as semiconducting, an out-of-plane piezoelectric polarization, direct band gap and robust Rashba effects which make them more attractive in modern nanomaterials science, engineering, renewable energy technology, high-end electronics and optoelectronic devices.[27,29,30] Lately, 2D Janus MoSSe material has been extensively studied providing great promise for their application in sensors, actuators, electrochemistry, photocatalysis, and other electromechanical devices.[27]

Recently, Lu et al. experimentally synthesized 2D monolayer Janus MoSSe TMD using chemical vapor deposition (CVD) technique from $MoS_2$ monolayers by substituting the top one S atom layers by Se atom layers.[25] It was experimentally observed that the Janus TMDs have shown hexagonal lattice structures but they differ in their electronic properties in comparison to symmetrically grown ordinary TMDs with their shattered out of plane structural symmetry.[31] This property indicates that the Janus TMDs may be an exceptional catalyst for the evolution of hydrogen. 2D monolayer Janus MoSSe TMD is a hexagonal 2D layer structure with six-fold Mo atoms and three-fold S and Se atoms by breaking the plane of symmetry.[32] Later on, Konkena et al. synthesized MoSSe nanosheets with a chemical vapor transport (CVT) technique using Mo, S, and Se precursor powders in precise proportions in a quartz tube at 800ºC temperature and normal atmospheric conditions.[33] This MoSSe was exfoliated into nanosheets and assembled with graphene oxide to construct a hybrid electrocatalyst for HER mechanism with the Tafel slope (*m*) about 51.0 mV.dec$^{-1}$ at 5.0 mA.cm$^{-2}$ current density.[33] Very recently, Pang et. al. theoretically studied that the Janus group-III chalcogenide monolayers can provide as a suitable substrate for silicene which might be applicable for nanodevices and optoelectronic devices.[34] It was experimentally observed that the electronic band gap (**E$_g$**) changes according to the direction of the applied external electric field. Li et al. computationally investigated the optical and electronic properties of the 2D Janus MoSSe material, and they reported that the electronic bandgap of the 2D Janus MoSSe was 1.47 – 2.69 eV computed by the PBE+SOC and G0W0+SOC methods.[27] Er et al. theoretically and computationally studied the properties of various 2D Janus TMDs and they found that the 2D Janus TMDs (monolayer) are promising candidates for HER catalysts.[24] However, they did not explore the detailed reaction mechanism and $H_2$ evolution pathways by investigating the materials/electronic properties, HER mechanism, activation barriers, pathways, and reaction kinetics with thermodynamics in detailed. We hypothesized that the 2D monolayer Janus MoSSe TMD could be used as an excellent electrocatalyst for HER due to a reasonable design to trigger the "inert" in-plane Se-Mo-S.

Here, we computationally developed 2D Janus MoSSe monolayer structure material, and studied the electronic band structure, total density of states (DOS), position of the Fermi energy level ($E_F$) and electronic band gap ($E_g$) with a potential application in electrochemical water splitting reactions via HER. In this work, we employed first principles-based hybrid periodic Density Functional Theory (DFT)[35,36] and van der Waals (vdW) corrections (i.e., Grimmes' -D3 dispersion corrections) to calculate the equilibrium geometries, structures, band structure, electronic band gap ($E_g$) and total DOS to predict the electronic/material properties of the 2D monolayer Janus TMD MoSSe. We found that the 2D Janus MoSSe single layer is an excellent material for $H_2$ evolution with high catalytic performance. It was found that the exposed S-/Se-edge ($\bar{1}010$) and Mo-edge ($10\bar{1}0$) edges of the 2D monolayer Janus TMD MoSSe are catalytic active for HER, and the (001) basal planes of the S−Mo−Se tri-layer of the MoSSe JTMD are exposed surfaces. It has more active sites for the hydrogen adsorption on the catalytic surface due to the presence of more surface to volume ratio compared to pristine 2D TMDs. To develop the 2D monolayer Janus MoSSe, we took a hexagonal symmetric 2D monolayer $MoS_2$ structure with ABA layered stacking (where A is Sulphur and B is Molybdenum) and replaced the top layer of S in the $MoS_2$ by the Se atoms layer to form 2D MoSSe structure with ABC layered stacking (A-Selenium, B-Molybdenum, and C-Sulphur) as shown in Fig. 1. The 2D layer structure was used to explore the electronic properties, and a molecular cluster model system $Mo_{10}S_{12}Se_9$ (noted by [MoSSe]) of the 2D MoSSe JTMD has been considered to investigate the hydrogen evolution reaction mechanism, reaction barriers, kinetics and thermodynamics with reaction pathways by using first-principle based M06-L DFT[35,36] calculations as depicted in Fig. 1. The molecular cluster $Mo_{10}S_{12}Se_9$ describes the Mo-edge and S-/Se-edge of the 2D monolayer Janus MoSSe which is essential to investigate the HER mechanism. It has shown that the DFT calculation is the *workhorse* to compute the electronic properties, thermochemistry, reaction kinetics, reaction barriers, and the activation energy barriers at various active sites of TMD's to investigate HER activity.[12,27,34,37–42] We computationally found that the Janus 2D MoSSe has exceptional electrochemical performance and electrochemical parameters like the lower value of the Tafel slope (59.16 mV/dec), lower activation energy barriers (both the H$^{\bullet}$-migration and Heyrovsky reactions: 3.93 – 7.10 kcal.mol$^{-1}$) and a larger value of turn over frequency (TOF: 3.87 x 10$^7$ – 8.16 x 10$^9$ sec$^{-1}$).

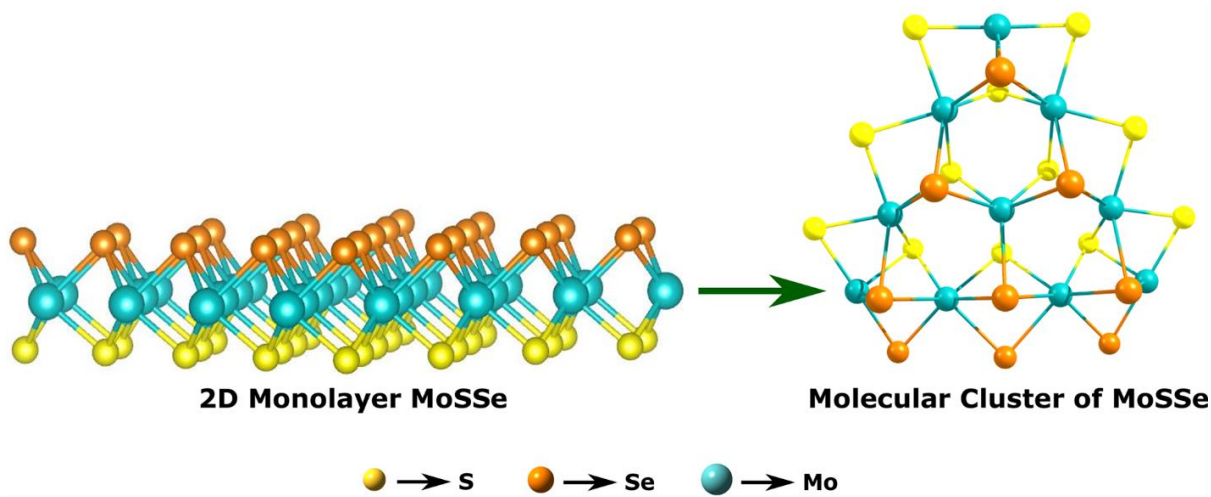

**Fig. 1:** Schematic presentation of the periodic 2D monolayer Janus MoSSe material (left side) and molecular cluster model system ($Mo_{10}S_{12}Se_9$) of the 2D MoSSe JTMD (right side).

## Methodology and Computational Details:

The progress in the science of advanced generations gives a clue to examine the interaction between molecules/atoms and catalysts in the submicroscopic scale. It is computationally very difficult to study the transition metal-containing molecules or complexes compared to the other row elements/atoms in the modern periodic table.[43] Recently developed plane-wave basis sets, Gaussian basis sets, Gaussian type of orbitals (GTO), pseudopotential and ultra-fast supercomputers made computations easier to explore the surface phenomenon of different catalysts for investigating various surface mechanisms.[44] Reaction barriers play a prominent role to understand the adsorption of an atom on the catalytic surface (adsorbent) to use it for various electrocatalytic applications. First-principles based DFT[35,36] methods are used to explore the electronic structure, properties, reaction pathways, kinetics, energetics, thermodynamics and mechanism for the $H_2$ evolution.[45,46]

## Computational Modeling of the 2D MoSSe JTMD:

## a) Periodic DFT Calculations of the 2D Monolayer MoSSe:

The equilibrium structures, geometries, and electronic properties of the 2D Janus MoSSe TMD material were calculated by the periodic hybrid HSE06 DFT method[47] as shown in Fig. 2a and Fig. 3. All these calculations (i.e. periodic 2D slab and electronic properties calculations) have been performed by using *ab initio-based* CRYSTAL17 suite code.[48,49] It should be noted here that the Gaussian type orbitals (GTOs) were used in the CRYSTAL17 code instead of the plane wave function which is implemented in the Vienna *Ab initio* Simulation Package (VASP) code.[50] The present method differs from the plane-wave based suite codes such as VASP, Wien2K, CP2K, CASTEP, Quantum Espresso, etc., however both the approaches reach to the comparable findings and outcomes *i.e.* both the VASP and CRYSTAL17 codes reach to the similar results.[36,37,41] It has been studied that the localized Gaussian basis set (GTO) codes are more intuitively appropriate for solving the Hartree−Fock (HF) part of the Schrödinger equations implemented in the hybrid density functionals menthods.[37,41,51–54] The periodic HSE06 DFT method has been implemented in the *ab initio*-based CRYSTAL17 suite program,[48] which effectively executes hybrid functional methods and facilitates standard application of the new methods to large-scale calculations of such materials with excellent performance, even with small-scale computing resources. The HSE06 method provides more accurate electronic band gap compared to the other DFT methods.[55] For prolonged van der Waals (vdW) interactions, Grimme's semi-empirical dispersion corrections (-D3) are implemented in the periodic HSE06 DFT calculations (in short DFT-D method). The quantum DFT-D (here HSE06-D3) method has been used for understanding the weak vdW interactions between the layers in 2D materials to obtain the accurate geometry, energy, electronic structure, electronic properties, etc. Triple-ζ valence with polarization (TZVP) quality Gaussian type basis sets have been applied for both the S and Se atoms for representing GTOs for all the layer structure calculations[56], and HAYWSC-311(d31)G Gaussian type basis sets with the Hay-Wadt type effective core potentials (ECPs) have been employed for the Mo atoms[57]. In general, the DFT-D methods offers a very good quality geometry of the 2D layered structure material after reducing the spin contamination effects such that it will not show any effect on the electronic structure and properties computations (i.e., electronic band gap, band structure and the total DOS).[2,5,58–60] The threshold employed for assessing the convergence of the energy, forces and electron density has been set to $10^{-7}$ a. u. for all the cases. The height of the unit cell is kept at 500 Å (which considers that there is no periodicity in the z-direction in the 2D layer structure model in the *ab initio-based* CRYSTAL17 code), i.e., the vacuum region in the z-axis was kept ~ 500 Å in these computations to accommodate the vacuum environment.[36,61]

The exposed surfaces of the (001) basal planes and the exposed edges of the ($\bar{1}010$) (S-/Se-edge) and ($10\bar{1}0$) (Mo-edge) boundaries of the 2D single layer Janus MoSSe TMD material are constructed here. The monolayer 2D MoSSe is terminated on these aforementioned edges boundaries during the 2D layer structure calculations as depicted in Fig. 2. It should be mentioned here that the (001) basal plane of the 2D MoSSe JTMD material with the Se-Mo-S configuration has exposed surface area at the edges of ($\bar{1}010$) S-/Se-edge and Mo-edge ($10\bar{1}0$) which are highlighted by two horizontal dashed lines shown in Fig. 2a. These two horizontal dashed lines imply the ends along the Se-/S-edge ($\bar{1}010$) and the Mo-edge ($10\bar{1}0$) as displayed in Fig. 2a. A recent experimental study on 2D TMDs materials proves this examination by evaluating the activities between the basal plane and edges.[12,62] It is expected that the electrochemistry of the edges of 3D bulk structure MoSSe to be analogous to that of a 2D monolayer MoSSe material. The Mo-/Se-/S-edges in the 2D Janus MoSSe layer show the activities for electrochemical HER, therefore, the 2D monolayer Janus MoSSe material is adequate to comprehend the electrochemistry of $H_2$ evolution reaction at the edges of this JTMD. It has been shown that the edges of the TMDs are active sites in the basal planes which are solely responsible for the electrochemical HER.[12,21] The optimized 2D MoSSe JTMD was visualized in VESTA, a visualization software that is used for its illustrations and analysis.[63]

Monkhorst k-mesh grids were used for computing the 2D electronic layer structure, geometry, total density of states (DOS), and electronic band structures calculations as shown in Fig. 3.[64] 20 x 20 x 1 k-mesh grids with a resolution $2\pi \times \frac{1}{60}$ Å$^{-1}$ were taken account to obtain both the electronic structure and properties computations of the 2D monolayer MoSSe JTMD, and the shrinking factor was about 65 during all the calculations. The total eight numbers of electronic bands were calculated around the Fermi energy level ($E_F$) in the first Brillouin zone along a high symmetric *Γ-M-K-Γ* direction which is consistent with the symmetry of the 2D MoSSe material as depicted in Fig. 3b. Both the electronic band structures and DOS were calculated with respect to the vacuum i.e., the electrostatic potential effects were added in the present calculations. All the atomic orbitals of the Mo, S, and Se atoms were employed to calculate and draw the entire orbitals electron density of states of the pristine 2D MoSSe JTMD as shown in Fig. 3c. DFT-D3 method with the Gaussian type basis sets provides comparable results to the highest-level $G_oW_o$ calculations (which are laborious and computationally extremely high-cost) available for significantly less cost.[36,61] Thus, the spin-orbit coupling effects have not taken into consideration while determining these band structure and total DOS (i.e., electronic properties) of the 2D monolayer MoSSe JTMD material.

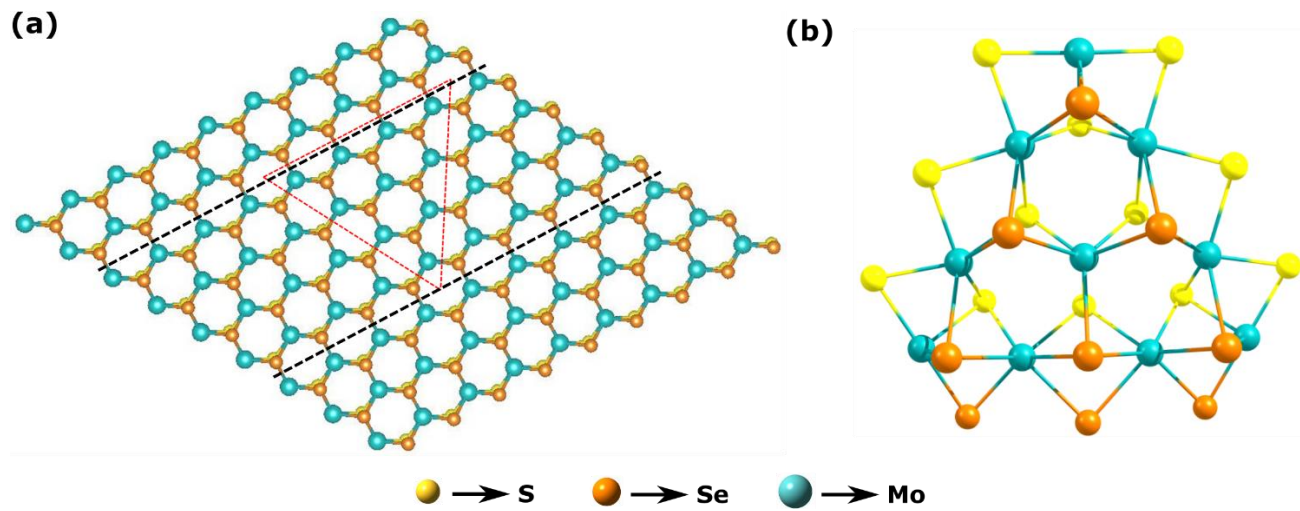

**Fig. 2:** (a) Top view of the 2D Janus monolayer MoSSe material is shown here. The two horizontal dashed lines reveal the ends along the ($\bar{1}010$) Se-/S-edge and ($10\bar{1}0$) Mo-edge of this JTMD. The triangle represents the ends for Se-/S-edge and Mo-edge finite molecular clusters. (b) Optimized triangular Mo-edge (with S-/Se-edge) cluster with stoichiometry of the 2D monolayer MoSSe in form of molecular cluster $Mo_{10}S_{12}Se_9$ (noted by [MoSSe]) is demonstrated here.

## b) Molecular Cluster Modelling and Non-periodic DFT Calculations:

A finite cluster model system has been constructed by considering the periodic structure of the 2D monolayer MoSSe JTMD material which is highlighted by a triangle as shown in Fig. 2. Two horizontal dashed lines indicate terminations along the ($10\bar{1}0$) Mo-edges and ($\bar{1}010$) S/Se-edges. The triangle represents the terminations for Mo-edge and S/Se-edge clusters and the dangling bonds in the finite cluster have been set by considering the same tringle as shown in Fig. 2. Each Mo atom in the basal plane (001) of the finite molecular cluster model has oxidation state of +4 and is bonded with the three numbers of the Se atoms at the upper plane and three numbers of the S atoms at the lower plane of Mo which gives a contribution of 4/6 = 2/3 electrons towards each Mo-Se and Mo-S bonding resulting a stabilized structure. The same can be understood with the oxidation state of the Se atoms in the basal plane. Selenium (Se) has -2 oxidation state and bonding with 3 Mo atoms results a contribution of 2/3 electrons towards each Mo-Se bond. Similarly, the edges of the periodic molecular cluster ($00\bar{1}0$) is being stabilized with the 2 local electron Mo-Se bonds (as well as Mo-S bonds) having a single electron contribution towards 4 Mo-Se bonding in the basal plane as shown in the Fig. 2. This 14/3 {i.e., (2×1) + [4× (2/3)]} contribution of electrons towards the Mo-Se bonding of the edge Mo atom is satisfied with the *d2* configuration of one Mo atom and *d1* configuration of two Mo atoms at the edges. This configuration leads the molecular system with the periodicity of 3 which results the achievement of a stabilized molecular cluster model having three edges without any unsatisfied valency. Thus, we considered a molecular cluster $Mo_{10}S_{12}Se_9$ model system (noted by [MoSSe]) to represent both the S- and Se-terminated Mo-edges on the surfaces of 2D monolayer MoSSe JTMD and this molecular model system is good enough to explain the HER process.

Fig. 2a demonstrates how the finite non-periodic molecular cluster model system has been constructed from the 2D periodic Janus MoSSe material to explain the exposed edges. Fig. 2b shows the equilibrium geometry of the molecular cluster model $Mo_{10}S_{12}Se_9$ (noted by [MoSSe] in the reaction) obtained by the M06-L DFT method. To investigate the performance of the 2D Janus MoSSe material for HER, the non-periodic finite molecular cluster model $Mo_{10}S_{12}Se_9$ system (as shown in Fig. 2b) has been considered here, and the M06-L[65,66] DFT method has been applied to investigate the reaction pathways, kinetics, barriers, and mechanism. This M06-L DFT method is a technique used for energetics, equilibrium structures, thermochemistry, and frequency calculations of the molecular cluster structures, and it has been found that the M06-L method provides excellent structures, geometries, reliable reaction energies and activation barriers for organometallic chemical reaction systems.[12,65,66] The 6-31+G** Gaussian types of atomic basis sets have been used for the light elements (such as O, H, as well as S atoms),[67] and the LANL2DZ Gaussian atomic basis sets with the relativistic effective core potentials (ECPs) have been applied for both the Se and Mo atoms. The transition structures or saddle points (both the H•-migration and Heyrovsky reaction steps) were computed to find the reaction barriers by confirming one imaginary frequency, modes of vibration and intrinsic reaction coordinate (IRC) calculations.[59,68] A transition state (TS) is a first order saddle point on a potential energy surface (PES) and it is a very short-lived configuration of atoms at a local energy maximum in a reaction-energy diagram also known as reaction coordinate. The transition state of a chemical reaction is a particular configuration (i.e. short-lived structure or geometrical configuration) along the reaction coordinate. A TS has partial bonds, an extremely short lifetime (measured in femtoseconds), and cannot be isolated. It is defined as the state corresponding to the highest potential energy along this reaction coordinate in PES. In other words, TS is an imaginary state between reactant and product. The vibrational spectrum of a transition state is characterized by one imaginary frequency (implying a negative force constant), which means that in one direction in nuclear configuration space the energy has a maximum, while in all other (orthogonal) directions the energy is a minimum. In order to verify if a stationary point is a transition state, a vibrational frequency calculation is performed at the same computational level as it has done in the geometry optimization. The normal mode corresponding to the imaginary frequency in the transition state usually reflects the change in geometry in going from reactants to products.[69] IRC can be defined as the minimum energy reaction pathway (MERP) in mass-weighted cartesian coordinates between the transition state of a reaction and its reactants and products. It can be thought of as the path that the molecule takes moving down the product and reactant valleys with zero kinetic energy.[59,68] IRC can be defined as the MERP in mass-weighted cartesian coordinates between the transition state of a reaction and its reactants and products. It can be thought of as the path that the molecule takes moving down the product and reactant valleys with zero kinetic energy. IRC approach has been used extensively in quantum chemical analysis and prediction of the mechanism of chemical reactions. The IRC gives a unique connection from a given transition structure to local minima of the reactant and product sides. This allows for easy understanding of complicated multistep mechanisms as a set of simple elementary reaction steps.[21,59,68]

Thermodynamic calculations and harmonic vibrational analysis were performed to obtain the thermodynamic potentials such as relative Gibb's free energy ($\Delta G$) and enthalpy ($\Delta H$). The imaginary frequencies obtained during frequency optimization were carefully eliminated to confirm the stable structure indicating local minima in the potential energy surfaces (PESs). To compute the solvation effects on the reaction barriers, polarizable continuum model (PCM) calculations have been performed by using water as a solvent with dielectric constant of 80.13. The PCM was employed for all the theoretical calculations to describe the solvation effects in the M06-L DFT computations, and three $H_2O$ molecules with one $H_3O^+$ have been combined unambiguously for the Heyrovsky reaction step. PCM is one of the best model to consider the solvation effects, and it is a commonly used method in computational quantum chemistry to model solvation effects.[12,40,51] The PCM represents one of the most successful examples among continuum solvation models, and it has been used to explain the solvation effects in various chemical reactions.[12,40,51] All the non-periodic molecular structure calculations, intermediates, transition states in the HER pathways and reaction barriers have been accomplished by using *ab-initio* based Gaussian 16 suite code.[70]

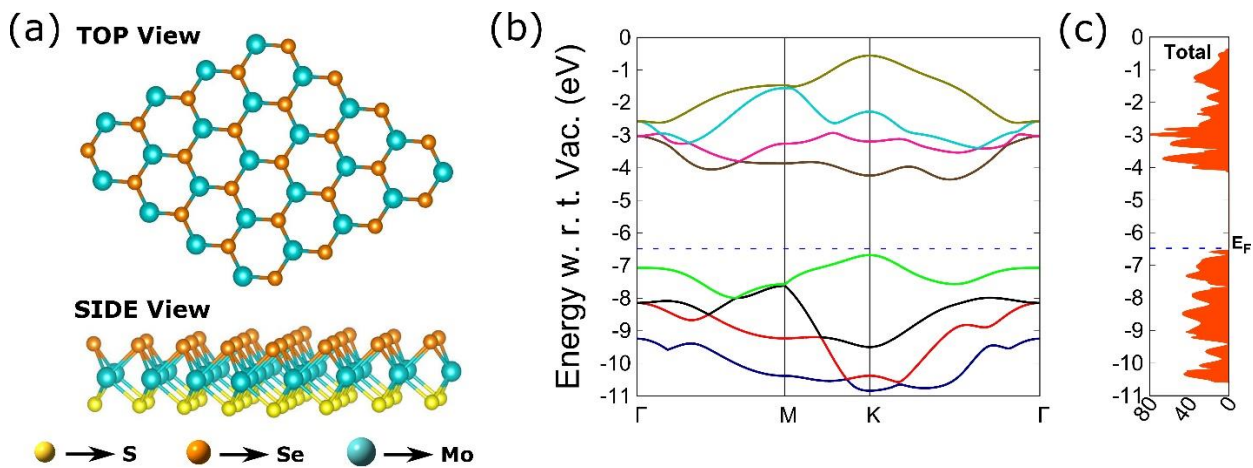

**Fig. 3:** (a) Top view and side view of the 2D monolayer MoSSe Janus TMD material; (b) Electronic band structure of the 2D single layer MoSSe Janus TMD; (c) Total density of states (DOS) of the 2D monolayer MoSSe Janus TMD material are displayed here.

**c) Validation of the Cluster Model:** For the validation of the molecular cluster model system $Mo_{10}S_{12}Se_9$ which has the same chemical properties as the 2D monolayer periodic slab Mo-edges of the MoSSe JTMD, we have calculated the hydrogen adsorption energies for both molecular cluster model and periodic 2D monolayer slab both under the vacuum conditions. The calculated values of hydrogen adsorption energies for both the systems are almost equal. Figure 4 shows that the hydrogen binding energies calculated by using cluster and periodic 2D slab only differ by 0.2 eV and 0.1 eV for the two stoichiometries. Thus, it can be mentioned here that the molecular cluster model has the same chemical properties compared with the periodic 2D Janus MoSSe.

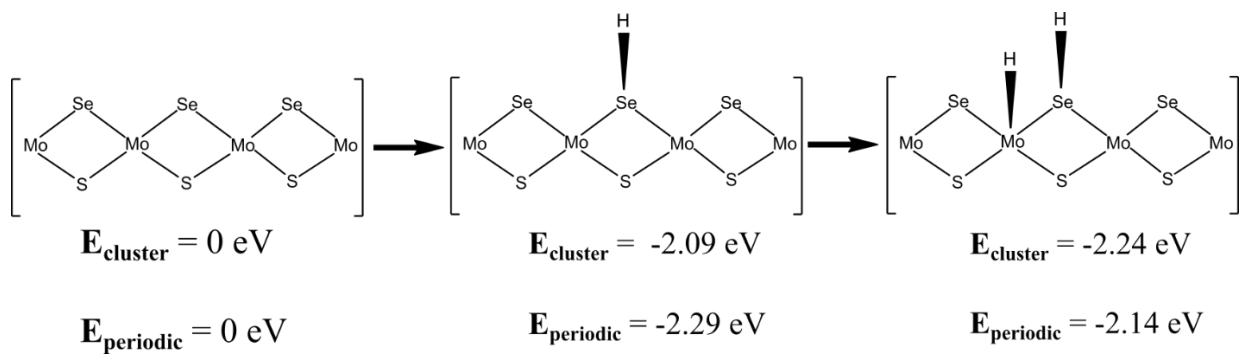

**Fig. 4:** Hydrogen adsorption energies on the Mo-edges of the 2D monolayer MoSSe JTMD. $E_{cluster}$ represents the relative electronic energy during hydrogen adsorption considering the molecular cluster model system and $E_{periodic}$ is the relative electronic energy obtained from the periodic 2D layer calculations.

## Theoretical Calculations and Equations:

The electrocatalytic performances of the 2D monolayer MoSSe Janus have been characterized by the computations of the changes of Gibbs free energy (ΔG) for $H_2$ adsorption on the ($\bar{1}010$) S-/Se-edges and ($10\bar{1}0$) Mo-edges of the MoSSe JTMD. The changes of free energy (ΔG), enthalpy (ΔH), and electronic energy (ΔE) in both the gas phase and solvent phase of the HER have been calculated by the following equations.

The value of relative Gibbs free energy (ΔG) (i.e., free energy changes) of [MoSSe]¯ is calculated by the subsequent equation below:

$$\Delta G = G_{[MoSSe]^-} - G_{[MoSSe]} - G_e \quad \ldots\ldots\ldots\ldots(1)$$

Where $G_{[MoSSe]^-}$, $G_{[MoSSe]}$ and $G_e$ are the free energy of [MoSSe]¯, [MoSSe] and free electron (e¯), respectively, calculated by the M06-L DFT method. The Gibbs's free energy ($G_e$) of an e¯ has been calculated at the standard hydrogen electrode (SHE) conditions where the $H^+$ and e¯ are in equilibrium with 1 atm $H_2$ at pH = 0. It is the difference between the free energies of 1/2$H_2$ and $H^+$ followed by previous works.[12,21]

The change of hydrogen adsorbed Gibbs free energy (ΔG_H) is calculated using the equation below

$$\Delta G_H = \Delta E_H + \Delta E_{ZPE} - T\Delta S \quad \ldots\ldots\ldots\ldots(2)$$

Where $\Delta E_H$ is the change of electronic energy of hydrogen adsorbed, $\Delta E_{ZPE}$ is a change of zero-point vibrational energy (ZPE) during hydrogen adsorbed, $\Delta S$ is the change in entropy.

$$\Delta E_H = E_{Janus+H} - E_{Janus} - E_{1/2H_2} \quad \ldots\ldots\ldots\ldots (3)$$

where $E_{Janus+H}$ is the one H atom adsorbed electronic energy of the JTMD (here 2D MoSSe), $E_{Janus}$ is the electronic energy of the pristine JTMD, and $E_{1/2H_2}$ is the energy of half of the molecular hydrogen ($H_2$).

$$\Delta H_H = H_{Janus+H} - H_{Janus} - H_{1/2H_2} \quad \ldots\ldots\ldots\ldots (4)$$

where $H_{Janus+H}$ is the enthalpy of the JTMD with one H atom adsorbed on its surface, $H_{Janus}$ is the enthalpy of the pristine TMD and $H_{1/2H_2}$ is the enthalpy of half of the $H_2$. We have used a similar formula to calculate the reaction barriers and relative energies in the HER. In the present computations, we have considered the SHE conditions. The change of Gibbs free energy of an electron at SHE condition was computed by considering the difference between the free energies of a $H^+$ (proton) as well as half of the $H_2$ molecule when pH = 0.

The calculation of the Gibbs free energy (G) of one hydrogen ($H_2$) molecule at 1 atm and 298.15 K was shown in the above section followed by Huang et al.[21] and Lie et al.[12] The Gibbs free energy (G) of a proton at 1 M in the solvent (here water) is about -270.3 kcal.mol$^{-1}$.[12] The absolute value of Gibbs free energy (G) of a $H^+$ in water solvent phase has been computed by considering its gas-phase value −6.3 kcal.mol$^{-1}$ plus the empirical hydration energy −264.0 kcal.mol$^{-1}$ followed by previous work Huang et al.[12] and Tissandier et. al.[71] The chemical potential ($\mu_e$) of $e^-$ and $H^+$ away from the SHE conditions are then calculated as:[12]

$$\mu_H(pH) = \mu_H(pH = 0) - 1.36 \times pH$$

And

$\mu_e(E) = \mu_e(SHE) - 23.06 \times E$ and the unit of both the cases are expressed in kcal.mol$^{-1}$.[12] The changes of free energies of all the non-periodic finite molecular clusters involved in the subject reaction with various numbers of $e^-$ and $H^+$ could be examined and estimated for the rest of the calculations by using the same M06-L DFT method by considering these reference values. The energies of all the systems such as [MoSSe], [MoSSeH$_{Se}$]$^+$, [MoSSe]$^-$, etc. considered in the molecular cluster model systems with the appropriate charge and multiplicity involved in the HER were computed by employing the equations 1-4.

The Turnover Frequency (TOF) at a specified temperature (T) can be obtained by applying the energetic span approximation i.e., transition state theory (TST); $TOF = \frac{k_B T}{h} \times \exp(-\frac{\Delta G}{RT})$; where k$_B$, R, h, and T are the Boltzmann constant, Universal gas constant, Plank's constant, and absolute temperature, respectively. ΔG is the change of free energy during the transition states formation in HER. The Tafel slope (*m*) has been computed by using the formula *m* = 2.3RT/nF; where F is Faraday constant, and n is number of electrons involved in the subject reaction. Tafel slope is an inverse measure of how strongly the reaction rate responds to changes in potential. It is used to evaluate the rate determining steps during the HER generally assume extreme coverage of the adsorbed species.

## Results and Discussions:

The equilibrium electronic structure of the 2D Janus MoSSe monolayer material (one-unit cell) calculated by the HSE06-D3 DFT technique is displayed in Fig. 3a with the electronic properties i.e., band structure and total DOS as shown in Fig. 3b-c, respectively. It should be noted here that the 2D monolayer Janus MoSSe material shows in-plane asymmetry with two trigonal prismatic lattices. The 2D MoSSe JTMD presents mirror symmetry with respect to the Mo plane breaks. The hexagonal Mo plane is squeezed in between Se and S layers with unequal distance.[27] It was computationally found that the 2D monolayer MoSSe JTMD material has hexagonal *P$_3$m$_1$* layer symmetry (2D plane layer group no. 69) with the lattice constants a = b = 3.256 Å, and interfacial angles α = β = 90° and γ = 120° obtained by the DFT-D (HSE06-D3) method as reported in Table 1 which reproduce the earlier experimental observation and data.[72] The lattice constants are also in accord with the previous computed results within 0.004 Å.[39] These parameters lie in between the values of the lattice constants and interfacial angles of both the pristine 2D monolayer MoS$_2$ and MoSe$_2$ TMDs, and it was found that the MoSSe has the structural phase 2H which is reasonable accord in previous studies.[27,39,71] The equilibrium Mo-Se and Mo-S bond lengths are about 2.489 Å and 2.380 Å, respectively, computed by the DFT-D method reported in Table 1, and these values agree with earlier reported experiment and theory.[27,72] Weak vdW interactions between atomic layers of 2D MoSSe structure show a significant effect on the average bond distances between the atoms present in each layer. These interactions increase the bond distances after taken into consideration in the structure calculations. The larger value of the bond distance between Mo and Se atoms compared to the Mo-S bond length is due to the large ionic radius value of the Se atom.[32] This variation in the bond distances and electronegativity of chalcogen anions (Se and S) present in the structure depicts that MoSSe is asymmetric (out of planar symmetry) i.e., because in the 2D MoSSe monolayer structure, Se and S atoms are placed in the top and bottom layers in between Mo atomic layer. This dissimilarity of the electronegativity of the chalcogens (here Se and S atoms) in the 2D monolayer MoSSe JTMD can help the electron transfer from the solvent to the reacting system which might have a potential application in HER.

**Table 1:** Equilibrium structural parameters (lattice constants and interfacial angles) of the 2D monolayer MoSSe JTMD material are reported here.

| Lattice parameters in Å | Interfacial angles in degree (º) | Space group And Symmetry | Average bond distance between atoms | |
|---|---|---|---|---|
| | | | Mo-S | Mo-Se |
| a = 3.256, b = 3.256 | A = 90º, β = 90º and γ = 120º | $P_3m_1$ Hexagonal | 2.308 | 2.489 |

The band structure of the 2D monolayer MoSSe Janus TMD has been plotted in high symmetric $\Gamma$-$M$-$K$-$\Gamma$ direction with respect to vacuum (w.r.t. vac.) as depicted in Fig. 3b. It was found that the Fermi energy level ($E_F$) is at -6.48 eV and the number of four conduction bands (CB) and valance bands (VB) around the $E_F$ level have been shown in the band structure calculations, and a direct electronic band gap is found at the *K*-point about 2.35 eV as displayed in Fig. 3b which agrees with the previous reported findings.[27,71] The direct electronic band gap (**$E_g$**) of the monolayer MoSSe JTMD is about 0.50 eV lower than the pristine 2D monolayer $MoS_2$ and $MoSe_2$ materials which illustrates that electron density is high in the case of asymmetrical TMDs like 2D Janus TMDs. This result agrees well with the previous computational results within 0.34 eV.[27] This lower value of electronic band gap (the energy gap between the valence bands (VB) and conduction bands (CB)) of the MoSSe JTMD helps to increase the conductivity of the structure and make it easy to transmit the electrons from the VB to the CB. The total density of states (DOS) is classified as the total number of the number of allowed electrons or holes states per volume at a given value of energy. In other words, DOS describes the number of states that are available in a system which is very essential for determining the carrier concentrations and energy distributions of carriers. The total density of states has been computed w.r.t. vac. and there are no energy band states in between -6.48 eV and -4.13 eV energy levels, and the density of states is higher at -3.0 eV energy level i.e., many energy states are accessible to be occupied as displayed in Fig. 3c. So, it means that the electron cloud is absent in the region where there is no energy band as shown in Fig. 3c.

The out of plane symmetry of the Janus MoSSe TMD structure causes charge imbalance due to the occurrence of different chalcogen atoms (here S and Se) which results in a potential in the vertical direction of the plane. This potential shows significant alterations in the electronic structures and properties of the Janus MoSSe TMD. Therefore, the coexistence of S and Se in the MoSSe introduces exposed edges and alters the atomic environment of the chalcogens and metals in this JTMD. It should be mentioned here that the electrocatalytic performance of the MoSSe for $H_2$ evolution matches up linearly with the total length of the exposed edges of the 2D monolayer MoSSe which can be found also in the case of 2D monolayer $MoS_2$ material.[21] The way to improve the electrocatalytic performance of the 2D Janus MoSSe material for the HER is by maximally exposing Se-, Mo- and S-edge sites using 2D monolayer MoSSe nanostructured material and the catalytically inert Se-/S-edge sites with additional exposure of active edge sites are also introduced by reducing the size of the material. In this case, the chalcogen atoms S and Se are completely responsible for changing the atomic atmosphere of the metal sites (here Mo) and chalcogen sites (here Se and S) of the 2D Janus MoSSe affecting the $H_2$ assimilation/absorption inside the (001) basal plane. However, these 2D electronic structure and properties computations and the results do not illuminate the exceptional performance of the 2D monolayer MoSSe Janus TMD material across the ordinary pristine 2D TMDs like $MoS_2$, $MoSe_2$, etc. for HER. To investigate the HER performance of the 2D Janus MoSSe material, a concrete study is required to examine and explain the electrochemical activities of the catalyst by investigating the detailed reactions mechanism, complete reaction pathways, kinetics, thermodynamics, and reaction barriers.

By employing the finite molecular model $Mo_{10}S_{12}Se_9$ system of the 2D monolayer MoSSe JTMD, we computed the changes of free energies ($\Delta G$) of the most likely intermediates with the transition states (TSs) to serve as a basis for describing the thermodynamics of HER which are relevant to the reaction process. The hydrogen adsorption free energy ($\Delta G_H$) has been computed by using the finite cluster molecular model $Mo_{10}S_{12}Se_9$ system with the M06-L DFT method. Estimating the hydrogen adsorption energy ($\Delta G_H$) on the surface of electrocatalyst is a useful descriptor to know its activity towards HER.[21,51,73] For an electrocatalyst to be efficiently active towards HER, it is necessary that the fast reactant supply and fast product delivery are required to be satisfied simultaneously. This requires the strong adsorption and strong desorption capabilities of hydrogen on the surface of catalyst which is hard to achieve at the same time. So, it is better to achieve a moderate hydrogen adsorption/desorption behavior balancing which offers an alternative way to enhance the HER performance of the catalysts. The hydrogen adsorption free energy ($\Delta G_H$) is regarded as a theoretical description of the HER activity. A positive value indicates low kinetics of adsorption of hydrogen molecules, whereas a negative value is low kinetics of release of hydrogen molecules. The ($\Delta G_H$) value should be close to thermo-neutral for the desired catalyst which should bind H neither too strongly nor too weakly.[21,51,73] The present DFT calculation reveals that the first H energetically prefers to make a bond to the Se atom by an amount of energy $\Delta G_H$ = 0.42 eV (which is equivalent to 9.65 kcal.mol$^{-1}$), whereas the value of $\Delta G_H$ is positive (+17.99 kcal.mol$^{-1}$ i.e. equivalent to +0.78 eV) when the H makes a bond with the Mo atom in the 2D MoSSe JTMD. In other words, the value of hydrogen adsorption energy $\Delta G_H$ at the metal side (here Mo) of the 2D MoSSe is about 0.78 eV (i.e. 17.99 kcal.mol$^{-1}$ i.e. the change of free energy or relative free energy) computed by the M06-L DFT method considering the $Mo_{10}S_{12}Se_9$ cluster model system. The value of $\Delta G_H$ is about 0.81 eV when the hydrogen adsorption happens at the S site in the 2D MoSSe cluster model system computed at the same level of theory. These results indicate that the first H energetically prefers to make a bond to the Se atom by an amount of energy 0.42 eV. The present computed results agree with the reported theoretical values of the 2D monolayer $MoS_2$ material wherever available.[21] Therefore, we can say that the direct hydrogen adsorption on the Mo-edge side is not thermodynamically favorable at the initial stage of the HER i.e., H would like to prefer to create the first bond with the Se atom in the 2D monolayer MoSSe JTMD.

In order to illustrate the $H_2$ evolution reaction mechanisms and the electrochemical catalytic activities as well as the performance of the 2D Janus MoSSe monolayer TMD material, we computationally explored the $H_2$ evolution reaction pathways considering two reaction steps; (i) Volmer reaction step where protons ($H^+$) interact with the electrons, and the hydrogen atoms get adsorbed on the active region of the catalyst (in short; **\* + H$^+$ + e$^-$ ⟶ H$^*_{ads}$**; where \* represents the active sites of the catalyst), and after the Volmer reaction, the H•-migrates from the S or Se (chalcogen site) to the Mo (metal site); and (ii) Heyrovsky reaction step where the $H_2$ has been formed by absorbing one solvated proton ($H^+$) from the solvent (here water) interacting with the absorbed H in the MoSSe. It has been considered here that the HER happens at the Se-terminated Mo-edges on the surfaces of 2D monolayer MoSSe JTMD, and the HER process at the S-terminated Mo-edges of the MoSSe JTMD has been also considered for comparing the performance of electrocatalytic activities. The detailed reaction pathway is shown in Fig. 5, and the thermodynamic computations have been performed at T = 298.15 K temperature. In addition, Volmer−Tafel mechanism has been also considered in the present study. The same molecular cluster model system $Mo_{10}S_{12}Se_9$ (noted by [MoSSe]) of the 2D Janus MoSSe has been considered for investigating the HER mechanism, reaction paths, reaction thermodynamics, kinetics, and reaction energy barriers of the H•-migration, Volmer−Tafel as well as Heyrovsky reaction steps. The equilibrium geometries, intermediates, and transition states (TSs) also known as 1st order saddle points have been computed by the same M06-L DFT method and they are displayed in Fig. 6a-i and 7. Various equilibrium bond lengths (average) of the pristine MoSSe, reactants, products, reaction intermediates, and transition states (TSs) appeared during the HER process are reported in Table 2. In order to investigate the HER in detailed, all the possible reaction steps have been considered. To proceed the HER, it is required to add protons ($H^+$) and electrons ($e^-$) simultaneously to the finite non-periodic molecular cluster model systems considered in the present study as depicted in Fig. 6-7. It should be mentioned here that it is beneficial to investigate all the equilibrium structures and stable geometries of the molecular clusters appeared during the HER process with the addition of extra electrons and protons to realize the changes of electronic energies and thermodynamical potentials i.e., ΔG and ΔH. These calculations will help us to find out the most thermodynamically favorable reaction pathways with the lowest reaction energy barriers in the $H_2$ evolution reaction process. In the HER steps, two exchange barriers (i.e., the largest and second-largest barriers) are important to assess the performance of the Janus MoSSe as an electrocatalyst.

**Table 2: Various equilibrium bond lengths of the different reaction intermediates and TSs with their charges. The unit of bond lengths is expressed in Å, and the charge is expressed in a.u.**

| Systems | Various equilibrium bond lengths of the different reaction intermediates and TSs with their charges | | | | | |
|---|---|---|---|---|---|---|
| | Equilibrium Bond Lengths (Å) | | | | Charge | No of Imaginary Frequencies |
| | Mo-Se | Mo-S | Se-H | Mo-H | | |
| [MoSSe] | 2.54 | 2.38 | - | - | 0 | 0 |
| [MoSSe]$^-$ | 2.57 | 2.32 | - | - | -1 | 0 |
| [MoSSe]H$_{Se}$ | 2.57 | 2.36 | 1.48 | - | 0 | 0 |
| [MoSSe]H$_{Se}^-$ | 2.56 | 2.40 | 1.48 | - | -1 | 0 |
| **TS1** | 2.58 | 2.40 | 1.64 | 2.00 | -1 | 1 (-146 cm$^{-1}$) |
| [MoSSe]H$_{Mo}^-$ | 2.52 | 2.38 | - | 1.72 | -1 | 0 |
| [MoSSe]H$_{Mo}$H$_{Se}$ | 2.56 | 2.37 | 1.48 | 1.72 | 0 | 0 |
| **TS2** | 2.56 | 2.41 | 1.48 | 1.85 | 1 | 1(-414 cm$^{-1}$) |
| [MoSSe]H$_{Se}^+$ | 2.54 | 2.38 | 1.48 | - | 1 | 0 |
| **TS3** | 2.55 | 2.43 | 1.51 | 1.74 | 0 | 1 (-322 cm$^{-1}$) |

The HER is initiated by the absorption of an electron in the 2D Janus Mo$_{10}$S$_{12}$Se$_9$ cluster (noted by [MoSSe]) at the SHE conditions leading to a negatively charged cluster [MoSSe]$^-$, and in the next step, hydride ion is absorbed on an energetically favorable Se site. This accompanies a hydride (H$^-$) shift to the reactive site of the catalyst, here it is the next adjoining metal atom (more specifically Mo) site. At first, one electron is transferred to the MoSSe formed [MoSSe]$^-$ with a change of favorable Gibbs free energy (ΔG) about -3.63 kcal mol$^{-1}$ at the SHE condition with E = 0 V and pH = 0 reported in Table 3. Therefore, the first reduction potential to form the [MoSSe]$^-$ was about -157 mV which indicates that a negatively charged [MoSSe]$^-$ cluster is thermodynamically stable in the solvated in water. In other words, the present DFT calculation indicates that the first reduction potential is -157 mV which leads to form the [MoSSe]$^-$ solvated in the water. It should be mentioned here that the equilibrium Mo-Se bond length has been slightly elongated by an amount 0.03 Å and the Mo-S bond length has been reduced by an amount 0.06 Å after the first electron reduction in the [MoSSe]$^-$. In the next step, one proton (H$^+$) is absorbed by the [MoSSe]$^-$ to form the [MoSSe]H$_{Se}$ and the proton is attached to the Se-edge side. The value of ΔG to adsorb the proton on the surface of the electrocatalyst [MoSSe]$^-$ is about -16.08 kcal mol$^{-1}$ with the change of enthalpy (ΔH) about -13.56 kcal mol$^{-1}$ as reported in Table 3. The equilibrium Mo-Se bond length remains the same in the [MoSSe]H$_{Se}$ system as the [MoSSe]$^-$, however, the equilibrium Mo-S bond length has been elongated by an amount 0.04 Å after the protonation. The equilibrium Se-H bond length Mo-Se is about 1.48 Å in the [MoSSe]H$_{Se}$ system computed at M06-L DFT method. The [MoSSe]H$_{Se}$$^-$ is formed due to the second electron transfer to the 2D Janus MoSSe material (i.e., [MoSSe]H$_{Se}$), and thus this HER is a two-electron transfer reaction. Both the equilibrium bond lengths Mo-Se and Mo-S have been found to be 2.56 Å and 2.40 Å, respectively, in the [MoSSe]H$_{Se}$$^-$ after second electron reduction, and the equilibrium Se-H bond length remains the same as [MoSSe]H$_{Se}$. In the next step of the reaction, the H$^\bullet$ migrates from the Se side to the Mo side which is called as the H$^\bullet$-migration step which occurs after the Volmer reaction. The Volmer reaction in the HER process at the (10$\bar{1}$0) Mo-edges occurs between the hydrogens on neighboring Molybdenum and Selenium atoms at the active sites of the 2D MoSSe JTMD material, and the adjacent H atom at the Se site moves towards the Mo site via a transition state (TS) which is called as H$^\bullet$-migration TS noted by (TS1). This TS1 has an imaginary frequency (~ -146 cm$^{-1}$) with mode of vibration towards the Se to Mo direction. The equilibrium structure of the TS1 is shown in Fig. 6e and the present DFT study reveals that the TS1 has equilibrium bond lengths about 1.64 Å for the Se-H (contrasted to a normal Se−H bond of 1.48 Å in the [MoSe]H$_{Se}$$^-$) and 2.00 Å for the Mo−H (contrasted to the equilibrium Mo−H bond of 1.72 Å in the [MoSe]H$_{Mo}$$^-$) as shown in Table . In other words, the Se-H bond length has been increased by an amount 0.16 Å during the H$^\bullet$-migration reaction (i.e., the H$^\bullet$ migrates from the Se site to Mo site in the HER process). The value of the change of enthalpy (ΔH) of the TS1 during the H$^\bullet$-migration reaction was found to be about 1.90 kcal mol$^{-1}$ and the reaction barrier (ΔG) is about 3.93 kcal mol$^{-1}$ computed in gas phase by DFT method as reported in Table 3. Therefore, it can be noted here that the reaction rate-determining step is the absorption of a proton and the formation of adsorbed hydrogen atom in the Volmer reaction mechanism during the reactions in gas phase. The present DFT computations reveal that this is the lowest H$^{\bullet}$-migration reaction barrier 3.93 kcal mol$^{-1}$ for the HER indicating that the 2D Janus MoSSe monolayer material is an excellent electrocatalyst to evolve H$_2$. It should be mentioned here that the present M06-L DFT calculations found that the first hydrogen strongly favors to create a bond to the Se atoms at the edges of selenium rather than the Mo atoms by an amount of 9.65 kcal mol$^{-1}$ energy (the change of free energy, $\Delta G$) and leads to a net binding energy relative to H$_2$ of 2.5 kcal mol$^{-1}$. After including the solvation effects (i.e., PCM calculations in water as a solvent), the corrections of entropy and zero-point vibrational energy (ZPE), the value of the $\Delta G$ due to the addition of hydrogen atom to the Se-edge is energetically favorable by an amount about 7.8 kcal mol$^{-1}$ computed by the M06-L DFT method. After the addition of one electron (e$^-$) to the cluster [MoSSe]H$_{Se}$, it reduces to [MoSSe]H$_{Se}^-$. When the system is negatively charged, H$^{\bullet}$ - migration occurs with a very small activation barrier of 3.93 kcal mol$^{-1}$ at a potential of 0.17 V. The direct hydrogen absorption or protonation in the metal side (here Mo) is energetically not favorable due to its high energy barrier of protonation. This study found that the reaction barrier energy $\Delta G$ is very high about 27.5 kcal mol$^{-1}$ when the direct protonation occurs at the Mo-edges, nevertheless, the value of $\Delta G$ at the Se-edges is reduced to -16.08 kcal mol$^{-1}$ with the change of enthalpy ($\Delta H$) about -13.56 kcal mol$^{-1}$ which is more thermodynamically favorable compared to the first case. Thus, the hydrogen absorbs at the Se site first, then, the H-atom migrates from the Se sites to the metal sites (here Mo atoms) with a very small reaction barrier ($\Delta G$) about 3.93 kcal mol$^{-1}$ known as H$^{\bullet}$-migration reaction barrier, and finally reacts with a proton from solution to form H$_2$. These multiple steps lower the reaction barrier for the whole process and make it energetically more favorable and suitable for further reaction.

In the next step of HER, H$_2$ has been formed due to the Heyrovsky reaction where three explicit water molecules with one hydronium (H$_3$O$^+$) cluster model system have been considered in the reaction to explain the H$_2$ formation as depicted in Fig. 6h. In other words, the last step of the HER is the H$_2$ formation which is known as Heyrovsky reaction mechanism as depicted in Fig. 5. In this step, the hydride H$^-$ (which has been migrated to the Mo-edges) reacts with a solvated proton (H$^+$) nearby explicit one H$_3$O$^+$ and three water molecules as shown in Fig. 6h. The optimized Heyrovsky transition state (TS2) obtained in the Heyrovsky reaction step is shown in Fig. 6h, and this TS2 corresponds to the formation of $H_2$ in the presence of four explicit water molecules ($3H_2O$ + $H_3O^+$ molecules) by two-electron ($2e^-$) transfer mechanism of HER. The average equilibrium O-H bond length in the water cluster of the Heyrovsky TS2 is about 0.97 Å, and the equilibrium Mo-Se and Mo-S bond lengths during the TS2 formation are about 2.56 Å and 2.41 Å, respectively, computed by the M06-L DFT method. The equilibrium Mo-H bond length was about 1.85 Å and the H-H bond length was about 0.93 Å during the formation of the Heyrovsky TS2 as depicted in Fig. 6h. This H-H bond length ($r'_{HH}$) in the TS2 is about 0.17 Å higher than the equilibrium H-H bond length ($r_{HH}$) of a pure $H_2$ molecule computed at the same level of DFT method. The equilibrium O-$H^+$ (where the $H^+$ is creating a bond with the H-Mo in the [MoSSe]$H_{Mo}H_{Se}$ of the TS2) bond distance in the Heyrovsky TS is about 1.28 Å. The source of a proton ($H^+$) is the solvation of hydronium ($H_3O^+$) in this reaction step, and the rate-determining step of this mechanism requires (in the gas phase reaction) the proton which is used to react with the adsorbed $H^-$ to form $H_2$. The present DFT computation found that the Heyrovsky reaction barrier is about 5.61 kcal.mol$^{-1}$ in the gas phase as tabulated in Table 3, and these two barriers, $H^{•-}$migration and Heyrovsky reaction barriers, are indicated by blue and red dotted circles in Fig. 5 (i.e. largest, and second-largest barriers). The position of $H^•$ in the TS1 during the $H^{•-}$migration reaction step and $H_2$ formation in the Heyrovsky step (TS2) was highlighted by a red dotted color in Fig. 6e and 6h. These two reaction barriers are the lowest compared to the other pristine 2D monolayer $WS_2$ or $MoS_2$ materials and their alloys as reported in Table 4.[12] Therefore, the present DFT calculations reveal that the electrochemical $H_2$ evolution performance of the 2D monolayer MoSSe Janus TMD material is better than other kinds of TMD materials such as pristine TMDs, i.e., $WS_2$ or $MoS_2$ materials and their hybrid alloys as both the reactions barriers ($H^{•-}$ migration and Heyrovsky reaction steps) for the $H^•$-migration and $H_2$ formation are the lowest among the other materials and electrocatalysts. This reaction mechanism provides intuitions on why 2D monolayer MoSSe JTMD is a very good electrocatalyst for effective $H_2$ evolution reaction.

There is another possibility to form $H_2$ by Volmer−Tafel reaction mechanism, where two hydrogens, which are already absorbed at the Mo- and Se-edges next each other in the molecular cluster [MoSSe]$H_{Mo}H_{Se}$, chemically react to produce a $H_2$ molecule. As the 2D monolayer MoSSe JTMD material has Se, Mo and S atoms in a tri-layer form Se-Mo-S, the types Se−H + H−Se and Mo−H + H−Se have been considered in the present calculations to explain the Volmer−Tafel reaction mechanism. This transition state geometry i.e. Tafel TS (TS3) was found to have bond distances of 1.74 Å for Mo−H (compared to the equilibrium Mo−H bond of 1.67 Å), 1.57 Å for H−H (compared to a final H−H bond of 0.76 Å), and 1.51 Å for Se−H (compared to a normal S−H bond of 1.48 Å) obtained by the M06-L DFT method. The Tafel TS geometry is depicted in Fig. 7 (at the right-hand side), and this H-H bond length ($r'_{HH}$) in the TS3 is about 0.81 Å higher than the equilibrium H-H bond length ($r_{HH}$) of a natural $H_2$ calculated at the same level of theory. It should be mentioned here that we considered both the Se and Mo atoms edges as the reaction occurs on the Se-/Mo-edges. However, impeding the H atoms on two neighboring Se and Mo atoms to move toward a possible transition state (TS3) known as Volmer−Tafel TS shown in Fig. 7. In other words, the absorbed two H-atoms in the [MoSSe]$H_{Mo}H_{Se}$ reacts each other via a possible transition state TS3 which is called Volmer-Tafel TS. Consequently, the Volmer−Tafel reaction on the Mo-edge occurs between the hydrogens on the adjacent Se and Mo atoms in the MoSSe JTMD during HER. The harmonic vibrational calculations and eigen vector analysis have shown that the TS3 has a single imaginary frequency about -322 cm$^{-1}$ before going to form $H_2$ during the HER. The reaction barrier (ΔG) was found to be 8.52 kcal.mol$^{-1}$ above the preceding intermediate [MoSSe]$H_{Se}H_{Mo}$ computed by the M06-L DFT method. The changes of electronic energy (ΔE) and enthalpy (ΔH) of the TS3 in this Volmer−Tafel step were about 8.68 and 8.49 kcal.mol$^{-1}$ with respect to [MoSSe]$H_{Se}H_{Mo}$, respectively, as depicted in Table 2. Therefore, these calculations indicate that the Volmer−Tafel reaction step is thermodynamically less favorable compared to the Volmer-Heyrovsky step for the $H_2$ evolution.

A further examination has been carried out to investigate the electrocatalytic activities of the S-terminated Mo-edges of the 2D monolayer MoSSe material for HER considering the same cluster model system. The same $H_2$ evolution reaction path (as depicted in Fig. 5) has been considered by taking account the S-terminated Mo-edges of the 2D monolayer MoSSe material i.e., the HER takes place at the S-terminated Mo-edges of the [$MoSSe$] considering the same finite cluster model system $Mo_{10}S_{12}Se_9$. Equilibrium geometries of the important intermediates and transition states involved in the subject reaction; [$MoSSe$]$H_S^-$, [$MoSSe$]$H_SH_{Mo}$, TS4 (H$^•$-migration) and TS5 (Heyrovsky) are shown in Fig. 8 as they are playing a significant role to determine the performance of electrocatalytic activities for effective HER. The present DFT-D calculations found that the H$^•$-migration reaction barrier (ΔG) is about 10.08 kcal.mol$^{-1}$, and the changes of electronic energy (ΔE) and enthalpy (ΔH) are about 10.10 and 9.70 kcal.mol$^{-1}$, respectively, when the reaction occurs at the S-terminated Mo-edges of the 2D monolayer MoSSe material. In this case, the reaction barrier (i.e., the change of Gibbs free energy ΔG) of TS4 is about 6.15 kcal.mol$^{-1}$ higher than the previous case TS1 when the reaction happens at the Se-terminated Mo-edges of this 2D MoSSe JTMD material. Similarly, the Heyrovsky transition barrier has been computed at the same level of DFT theory at the S-terminated Mo-edges as depicted in Fig. 7d, and the values of TS5 barrier (ΔG) is about 14.05 kcal.mol$^{-1}$, and the values of ΔE and ΔH are about 13.22 and 12.58 kcal.mol$^{-1}$, respectively. The Heyrovsky's reaction barrier is about 8.44 kcal.mol$^{-1}$ higher than the same barrier of the TS2 when the HER happens at the Se-terminated Mo-edges. These calculations indicate that the HER process at the S-terminated Mo-edges of the 2D MoSSe JTMD is thermodynamically less favorable than the Se-terminated Mo-edges.

**Fig. 5:** Hydrogen evolution reaction (HER) mechanism with pathway occurs on the surfaces of 2D monolayer MoSSe JTMD as an electrocatalyst; H•-migration reaction step is highlighted by red ellipse with a barrier about 7.10 kcal.mol$^{-1}$ in the solvent phase (water) and H$_2$ formation indicates the Heyrovsky reaction step highlighted by blue ellipse with a barrier 4.72 kcal.mol$^{-1}$ in solvent phase. The HER happens at the Se-terminated Mo-edges on the surfaces of 2D monolayer MoSSe JTMD.

**Table 3:** Comparison of various energy values (ΔE, ΔH and ΔG) of the 2D MoSSe JTMD catalyst during the HER process in the Gas Phase calculations is tabulated here. The units of the change of electronic energy (ΔE), enthalpy (ΔH) and Gibbs free energy (ΔG) are expressed in kcal.mol$^{-1}$. The HER happens at the Se-terminated Mo-edges on the surfaces of 2D monolayer MoSSe JTMD.

| HER Reaction Steps | ΔE kcal.mol$^{-1}$ (Gas Phase) | ΔH kcal.mol$^{-1}$ (Gas Phase) | ΔG kcal.mol$^{-1}$ (Gas Phase) |
|---|---|---|---|
| [MoSSe] → [MoSSe]$^-$ | -3.63 | -3.45 | -3.62 |
| [MoSSe]$^-$ → [MoSSe]H$_{Se}$ | -14.14 | -13.56 | -16.08 |
| [MoSSeH$_{Se}$] → [MoSSe]H$_{Se}^-$ | -1.46 | -1.15 | -2.92 |
| [MoSSeH$_{Se}$]$^-$ → TS1 | 2.59 | 1.90 | 3.93 |
| TS1 → [MoSSe]H$_{Mo}^-$ | -2.73 | -3.28 | -2.01 |
| [MoSSe]H$_{Mo}^-$ → [MoSSe]H$_{Mo}$H$_{Se}$ | -21.25 | -20.98 | -21.01 |
| [MoSSe]H$_{Mo}$H$_{Se}$ → TS2 | 3.40 | 3.24 | 5.61 |
| TS2 → [MoSSe]H$_{Se}^+$ | -44.05 | -43.84 | -42.80 |
| [MoSSe]H$_{Mo}$H$_{Se}$ → TS3 | 8.68 | 8.49 | 8.52 |

**Table 4.** A summary of the electrochemical performance for HER of the 2D monolayer TMDs, transition metal phosphides (TMPs) (such as WS$_2$, MoS$_2$, W$_x$Mo$_{1-x}$S$_2$, MoP and MoSP) and the Janus TMD MoSSe materials with the values of the activation barriers, TOF and Tafel slope (*m*) for comparison are given below. The barriers are expressed in kcal.mol$^{-1}$.

| 2D TMDs/JTMDs | H$^-$ migration reaction barrier (kcal.mol$^{-1}$) | Heyrovsky reaction barrier (kcal.mol$^{-1}$) | Tafel Slope (*m*) (mV.dec$^{-1}$) | Turn Over Frequency (TOF) (sec$^{-1}$) | References |
|---|---|---|---|---|---|
| MoS$_2$ | 17.7-20.5 | 23.6-23.8 | 60 | 2.1x10$^{-5}$ | [12, 21] |
| WS$_2$ | 12.4-18.1 | 14.5-21.3 | 55-60 | 1.5x10$^{-3}$ | [12] |
| W$_x$Mo$_{1-x}$S$_2$ | 6.8-11.90 | 11.5-13.3 | 59 | 1.1x10$^3$ | [12] |
| MoP | 17.5 | 16.3 | 56-60 | 9.2x10$^{-1}$** | [49] |

| MoSP | 12.90-14.0 | 9.2-12.90 | 34 | 2.2x10³** | [49] |
| MoSSe* | 3.93-7.10 | 4.72-5.61 | 59.16 | 3.87 x 10⁷ | This work. |

*The HER happens at the Se-terminated Mo-edges on the surfaces of 2D monolayer MoSSe JTMD.

** Computed using the previous reported experimental results.

The reaction barrier energy and pathways have been computed in the solvent phase studying polarization continuum model (PCM) analysis, and the solvent phase computations have been performed by considering water as a solvent. It should be noted here that a dielectric constant value of 80.13 was used in the PCM calculations, which helps to explore the solvation effects in HER mechanism of the MoSSe JTMD. The H•−migration activation energy barrier (here $\Delta E$) for HER in the solvent phase is about 7.10 kcal mol$^{-1}$ which is the lowest compared to the pristine MoS$_2$ and WS$_2$ materials and their alloys (11.9 – 18.1 kcal mol$^{-1}$).[12] The activation energy barrier (here $\Delta E$) for the H$_2$ formation by the Heyrovsky mechanism on the surfaces of the 2D MoSSe is about 4.72 kcal mol$^{-1}$ in the solvent phase as shown in Table 5, which is also the lowest barrier compared to the pristine TMDs and their alloys as depicted in Table 4.[12] These lower values of both the activation energy barriers lead to the establishment of a sophisticated number of active sites by shrinking the value of electrochemical properties like Tafel slope (*m*) and increasing the turnover frequency (TOF) for an excellent catalytic activity to generate molecular hydrogen. A comparison has been drawn to predict the electrocatalytic HER efficiency of the pristine 2D monolayer MoSSe JTMD material with the other TMDs, their hybris alloys, and transition metal phosphide (TMP) MoP and S-doped MoP. The present study illustrates that the pristine 2D monolayer MoSSe JTMD material has shown the lowest reaction barriers (both the H•−migration and Heyrovsky) and the highest value of TOF compared to the pristine TMDs, their hybrid alloys, MoP and S-doped MoP (i.e. MoSP) as shown in Table 4. The Highest Occupied Molecular Orbitals and Lowest Unoccupied Molecular Orbitals (HOMO-LUMO) calculations on both the TS1 and TS2 have been performed to get a scientific insight into the electronic role in the HER mechanism. An explanation can be given from the HOMO-LUMO structures of both the TS1 and TS2 to support the lower activation barriers in the HER. The equilibrium HOMO-LUMO structures of both the TS1 (H•−migration TS) and TS2 (Heyrovsky TS) are depicted in Fig. 9. Accordingly, the Heyrovsky mechanism, in which the H atom bound to Mo site recombines to form H$_2$ is the one present on the Janus MoSSe. The electrochemical parameters Tafel slope and turn over frequency (TOF) were computed by the DFT method, and the values are 59.16 mV/dec and 2.15 x $10^9$ sec$^{-1}$. The computed TOF of the MoSSe JTMD is the highest compared to the ordinary TMDs and their alloys.[12] The DFT-D estimated Tafel slope (*m*) of this 2D MoSSe material is comparable to the TMDs alloys $W_xMo_{1-x}S_2$.[12] The formation of the $H_2$ molecule is confirmed by the overlapping of *s*-orbitals of the H-atoms connected to the Mo atoms in the MoSSe JTMD and *s*-orbitals of the hydrogen atoms in the water cluster.

**Table 5:** Reaction barrier energies and TOF obtained by the M06-L DFT method in computed the gas phase and solvent phase during the HER process at the Se-terminated Mo-edges on the surfaces of 2D monolayer MoSSe JTMD are tabulated here.

| System | ΔG kcal/mol (Gas phase) | ΔE kcal/mol (Solvent Phase) | Turn Over Frequency sec$^{-1}$ (Gas phase) | Turn Over Frequency sec$^{-1}$ (Solvent phase) |
|---|---|---|---|---|
| H$^-$ migration reaction step | 3.93 | 7.10 | 8.16 x $10^9$ | 3.87 x $10^7$ |
| Heyrovsky reaction step | 5.61 | 4.72 | 4.79 x $10^8$ | 2.15 x $10^9$ |

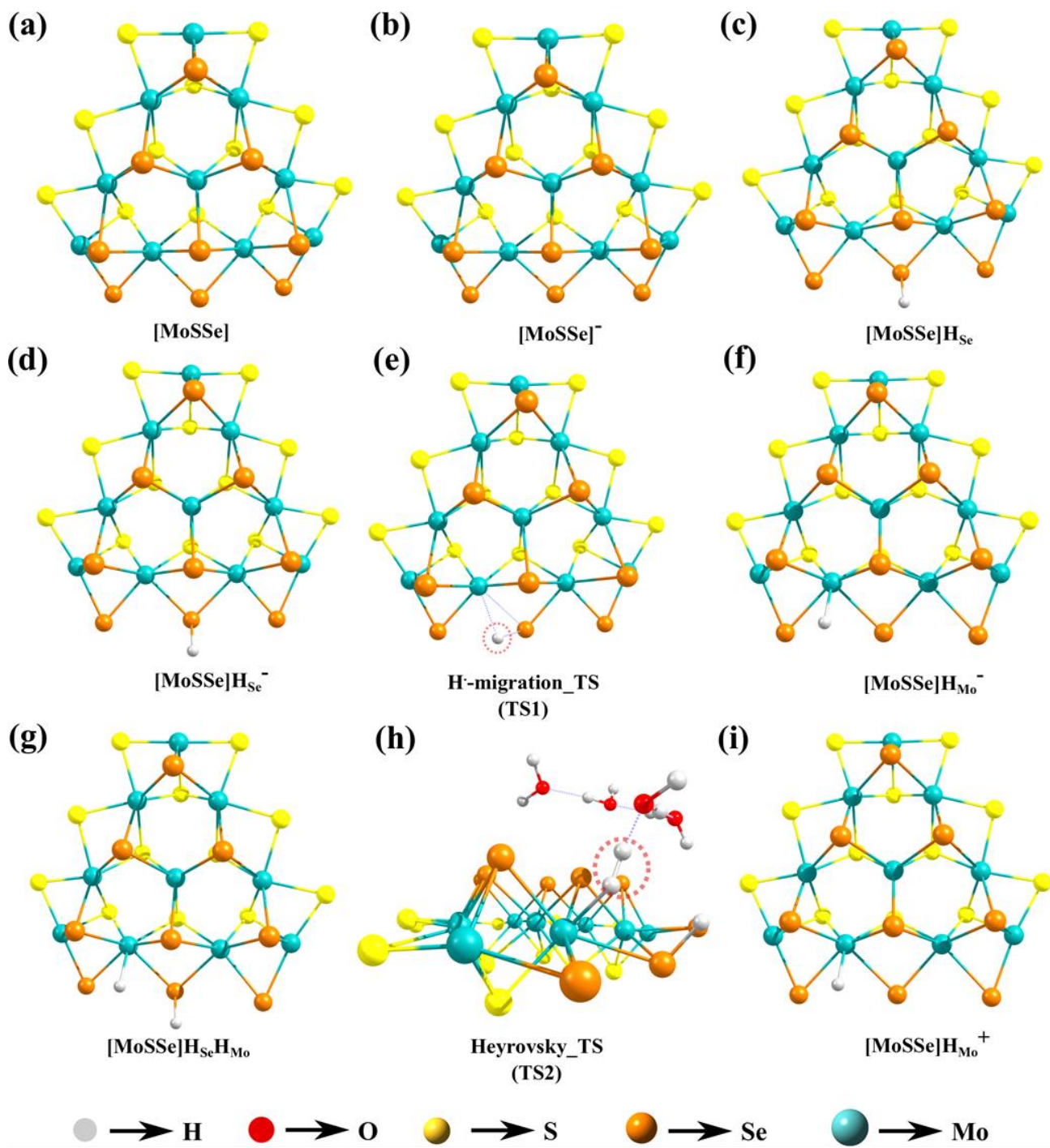

**Fig. 6:** Equilibrium geometries of (a) **[MoSSe]**, (b) **[MoSSe]⁻**, (c) **[MoSSe]H$_{Se}$**, (d) **[MoSSe]H$_{Se}^-$**, (e) H·-migration transition state i.e. **TS1**, (f) **[MoSSe]H$_{Mo}^-$**, (g) **[MoSSe]H$_{Se}$H$_{Mo}$**, (h) Heyrovsky's transition state i.e. **TS2** and (i) **[MoSSe]H$_{Mo}^+$** computed by the M06-L DFT method considering a molecular cluster model system Mo$_{10}$S$_{12}$Se$_9$ to represent 2D monolayer MoSSe JTMD are shown here. The HER happens at the Se-terminated Mo-edges on the surfaces of 2D monolayer MoSSe JTMD.

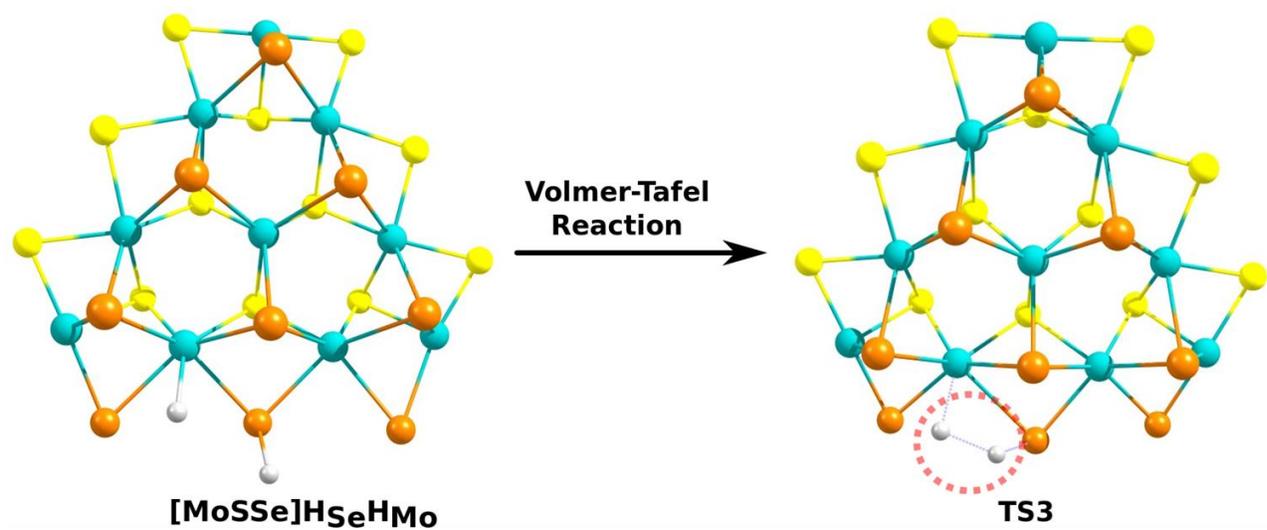

**Fig. 7:** Equilibrium structures of the **[MoSSe]$H_{Se}H_{Mo}$** and **TS3** occurred during the Volmer-Tafel reaction step are depicted here.

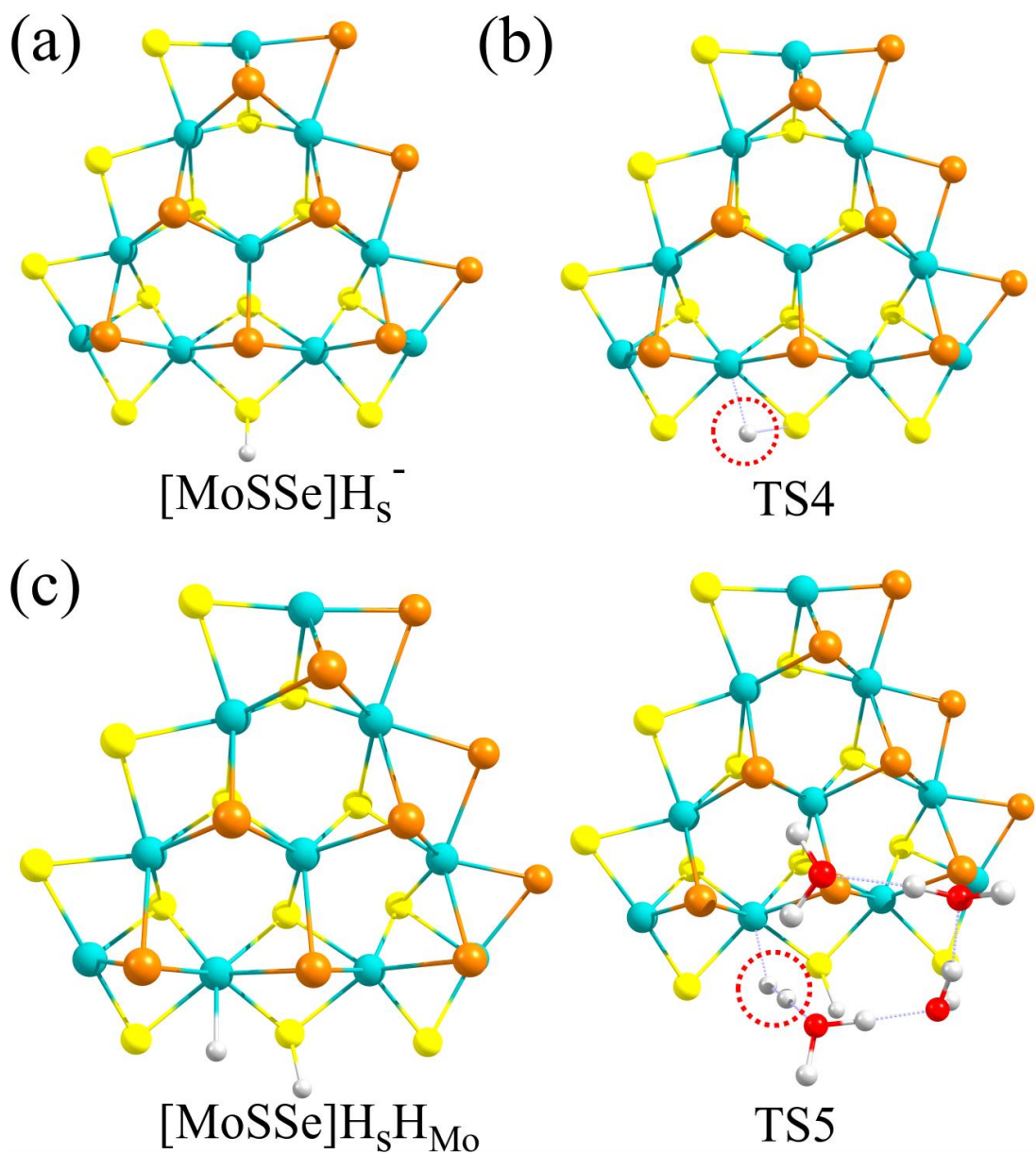

**Fig. 8:** Equilibrium geometries of the S-edges: (a) **[MoSSe]H$_S^-$**, (b) **TS4 (H˙-migration Transition State)**, (c) **[MoSSe]H$_S$H$_{Mo}$** and (d) **TS5 (Heyrovsky Transition State)** computed by the M06-L DFT method considering a molecular cluster model system Mo$_{10}$S$_{12}$Se$_9$ [noted by [MoSSe]] to represent 2D monolayer MoSSe JTMD are shown here. The HER happens at the S-terminated Mo-edges on the surfaces of 2D monolayer MoSSe JTMD.

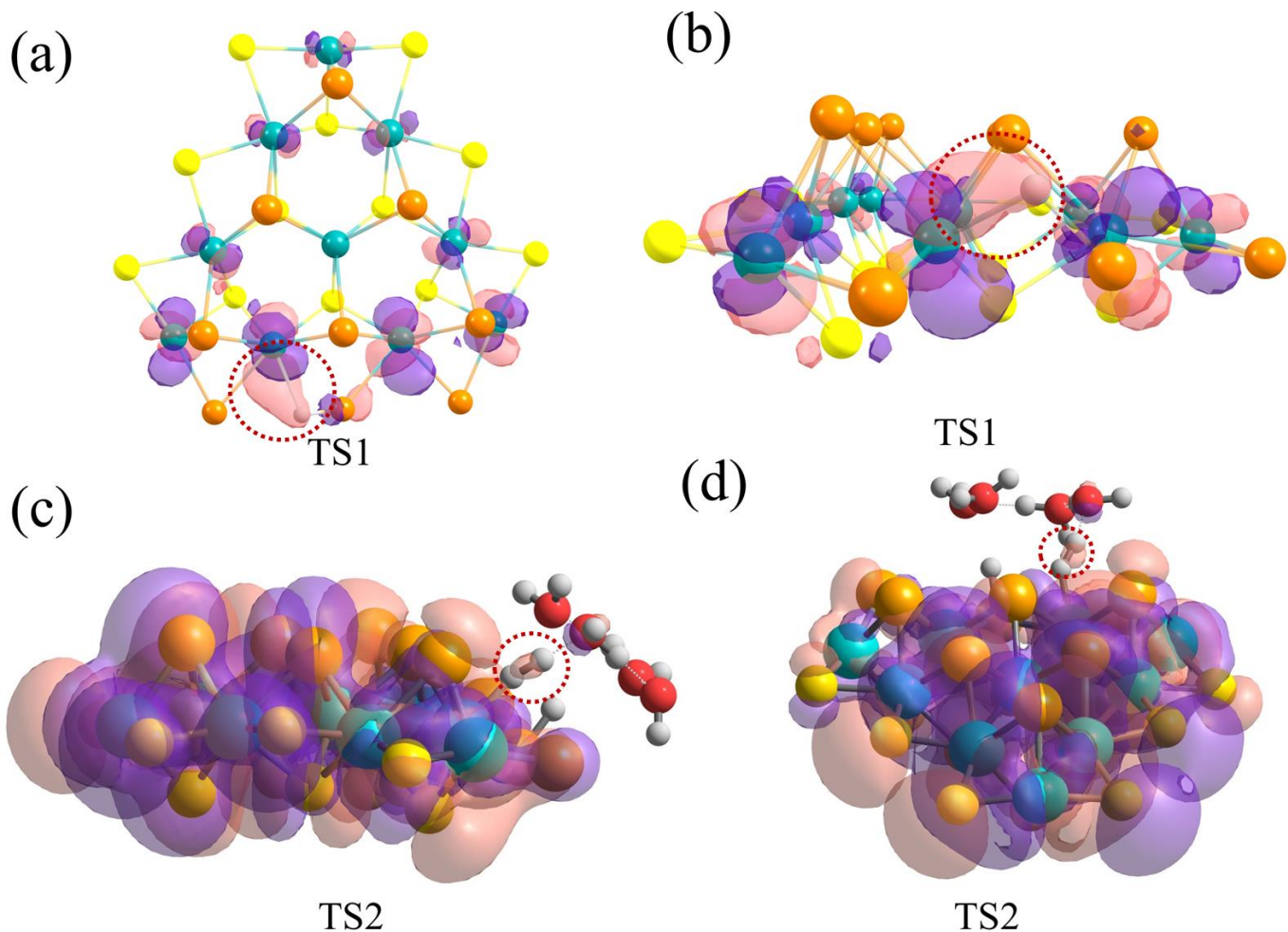

**Fig. 9:** The HOMO-LUMO of both the transition state structures TS1 (i.e. H˙-migration Transition State) and TS2 (Heyrovsky Transition State) occurred in the H$_2$ evolution reaction are shown for (a) Top and (b) side view of TS1; and (c) Top and (d) side view of TS2 during the HER on the exposed edge surfaces of the 2D MoSSe JTMD where the molecular orbitals involved in the H˙-migration and H$_2$ formation are highlighted by a red dotted circle. The HER happens at the Se-terminated Mo-edges on the surfaces of 2D monolayer MoSSe JTMD.

The present DFT computations revealed that both the reaction barriers (H˙-migration and Heyrovsky) for HER are the lowest among the ordinary 2D TMDs (such as MoS$_2$, WS$_2$, etc.) and these values are very close to the Pt(111) catalyst when the HER takes places at the Se-terminated Mo-edges on the surfaces of 2D monolayer MoSSe JTMD. To support our computational results and explain the HER activity of the 2D Janus MoSSe, we further extended the Natural Bond Orbital (NBO), the HOMO and the LUMO calculations of both the H•-migration and Heyrovsky transition state (noted by TS1 and TS2) which can explain the HER mechanism from the electronic charge as well as overlapping of the atomic/molecular orbitals. The HOMO and LUMO were obtained from the optimized transition states of both the H•-migration and Heyrovsky reaction step as shown in Fig. 9a-d, and the charge clouds were indicated by a red dotted circle between two hydrogen atoms in which H-migrates from the Se to Mo (in TS1) and the $H_2$ is formed (in TS2) due to the overlapping of the molecular orbitals. The NBO study delivers the most probable 'natural Lewis structure' picture of the wave function ($\varphi$), such that total information related to orbitals is selected mathematically to consider the maximum possible energy of electron density.[74] An advantage of this orbital is that it gives information about the intra-molecular and inter-molecular interactions (i.e. connections between bonds of atoms of the molecule). Donor acceptor interactions in NBO calculations are known from second order fock-matrix.[75] Chemical stability and reactivity of molecules can be understood from the HOMO and LUMO energies with the wave functions. The structural stability of a molecule is known from the energy difference between HOMO and LUMO energies.[76, 77] Electron donor unit and electron acceptor unit capabilities are known from HOMO and LUMO energies, respectively.

The total wave function of the HOMO and LUMO appeared at the equilibrium Heyrovsky transition structure TS2 is the linear combination of atomic orbitals (LCAO) of the Mo, S, H, O, and Se atoms obtained by the NBO calculations. Similarly, the total wave function of the HOMO-LUMO structure of the TS1 is the LACO of the Mo, S, Se and H atoms as shown in Fig. 9a-b. It should be mentioned here that the rate-determining step of the $H_2$ evolution reaction in gas phase is Heyrovsky reaction step as the barrier is 5.61 kcal.mol$^{-1}$. Whereas the H•-migration reaction step is the rate-determining step of HER in the solvent phase (i.e., water) as the reaction barrier during the H•- migration reaction is about 7.10 kcal.mol$^{-1}$ computed by the M06L DFT method. The HOMO structures of both the transition states TS1 and TS2 provide an intuition of the electronic role in both the H•-migration and Heyrovsky steps during the $H_2$ evolution. This study found that the TS1 are alleviated by the overlap of the *s*-orbital of the $H_2$, and the *d*-orbitals of the Mo metal atoms as displayed in Fig. 9a-d. The present computations shows that the Tafel slope is lower for the 2D Janus MoSSe due to the presence of this overlap of *s*-orbital of the hydrogen and the *d*- orbitals of Molybdenum atoms appeared in the HOMO and LUMO structure the transition states TS1 during H˙-migration reaction step. A better atomic orbital overlap of the *s*-orbitals of the hydrogen atom attached with the Mo in the MoSSe JTMD and the water cluster ($3H_2O + H_3O^+$) seemed in the HOMO-LUMO Heyrovsky's transition state TS2 during $H_2$ formation has been observed in Fig. 9c-d. This better overlap of the atomic orbitals during the $H_2$ creation in the Heyrovsky's TS2 reduces the reaction barrier. Therefore, it can say that this better stabilization (compared to the pristine $MoS_2$ and $WS_2$)[12] of the orbitals in the reaction limiting step TS1 for H˙-migration (solvent phase) and TS2 for $H_2$-formation (in the gas phase) is a key for reducing the Heyrovsky's reaction barrier, thus the overall catalysis indicating a better electrocatalytic performance for $H_2$ evolution. The electronic charged cloud represents the positive and negative part of the wave functions with the violet and pink color transparent cloud around Mo, S, Se, O, and H atoms as shown in Fig. 9. The electron cloud (noted by pink color) in between two H atoms in the Heyrovsky transition state structure highlighted by the red dotted circle is more in the case of pristine 2D MoSSe JTMD in comparison to the pristine 2D $MoS_2$ TMD structure in presence of four explicit water molecules in Heyrovsky step of $H_2$ evolution. Therefore, the Heyrovsky transition state TS2 of the rate-limiting reaction step in the gas phase calculation i.e., the Heyrovsky reaction step has been sustained by overlapping the atomic *s*-orbitals of the H atoms attached with the Mo atom and the adjacent hydronium ($H_3O^+$) in the explicit water cluster to form the $H_2$ molecules during HER. Similarly, the TS1 of the rate-limiting reaction step of the H˙-migration step (in the solvent phase calculation) was prolonged by the *d*-orbitals of the Mo atoms and the *s*-orbital of the H atom during the HER process as depicted in Fig. 9a-b. It should be noted here that these overlapping with better stabilization (compared to the pristine 2D monolayer $MoS_2$ and $WS_2$)[12] determines the reaction barriers and electrocatalytic performance of the 2D MoSSe JTMD for effective HER. This is the reason behind the 2D Janus MoSSe material which has shown an excellent catalytic activity for effective HER, and the electron transfer from the solvent to the catalysts is faster than ordinary pristine 2D TMDs due to different electronegativity of the chalcogen Se and S atoms in the 2D Janus MoSSe TMD.

In the Janus structure of the 2D MoSSe material, the breaking of structural symmetry provides a large inherent lattice stress and has substantial impacts in modifying the electronic band structures for characteristic physical, electronic, materials and chemical properties. The advent of the in-gap states and the change in the Fermi energy ($E_F$) level of the Janus structure of this 2D monolayer MoSSe JTMD material due to the structural asymmetry largely effect the hydrogen adsorption on the surfaces of it. It should be mentioned here that both the chalcogen atomic edges (Se and S) and metal edges (i.e., Mo-edge) of the 2D monolayer MoSSe JTMD material are catalytically active for the $H_2$ evolution reactions (as shown in our present computed results) leading to an excellent electrocatalytic performance compared to the pristine 2D $MoS_2$ TMD. The present study furnishes a unique computational technique to enhance the $H_2$ creation through the development of novel 2D MoSSe JTMD and electrochemical HER process on the edges of the Janus monolayer MoSSe material. It has been studied that the Janus structure's intrinsic dipole lead to considerable bending of bands for obtaining band edge positions comparable to redox potentials of water, which renders the 2D monolayer MoSSe JTMD as an auspicious electrocatalyst for gaining the maximum amount of hydrogen adsorption on its surface.[72] It should be mentioned here that the Janus asymmetry in the 2D monolayer MoSSe material is the origin of enhanced HER activity and it is minimizing both the $H^{\bullet-}$migration and Heyrovsky reaction barriers due to the atomic orbitals overlap ($s$-orbital of the hydrogen and $d$-orbitals of Mo atoms in $H^{\bullet-}$migration TS $s$-orbital of the hydrogen atoms in the case of Heyrovsky TS) during the TS formation in HER. Therefore, the present DFT results have shown the way of pavement to develop a high-performance and earth abundant asymmetric 2D Janus monolayer TMD materials for an efficient HER electrocatalysts. This study has shown an improved catalytic property of the JTMD toward highly efficient production of $H_2$ by providing a minimum energy reaction pathway forward for sustainable $H_2$ evolution from the 2D monolayer MoSSe Janus TMD driven catalysis.

## Conclusion:

In summary, we computationally developed 2D Janus MoSSe material by employing hybrid periodic HSE06-D3 DFT method and studied their electronic properties i.e., band structure and total density of states. The electronic property calculations showed that the asymmetric 2D monolayer MoSSe Janus TMD is a semi-conductor with a direct electronic band gap about 2.35 eV, which is lower than the pristine TMDs, and the total density of states calculations depict that electronic density has been increased in asymmetrical 2D Janus MoSSe. It has been found that the exposed surfaces with the Se-/S-edges ($\bar{1}010$) and Mo-edges ($10\bar{1}0$) of the 2D monolayer Janus TMD material are electrocatalytic active for HER. Modeling these active-edges of a monolayer 2D Janus MoSSe TMD material with a non-periodic finite $Mo_{10}S_{12}Se_9$ molecular cluster system, the present DFT study showed that the $H_2$ evolution occurs through the Volmer−Heyrovsky reaction mechanism including a $H_3O^+$ and an $e^-$ rich molybdenum hydride. In other words, the HER process on the exposed surfaces of the 2D MoSSe JTMD follows the most thermodynamically favorable Volmer−Heyrovsky reaction mechanism instead of Volmer-Tafel reaction mechanism. It was computationally found that the 2D Janus MoSSe monolayer material is an outstanding electrocatalyst for HER. It was observed that the overall chemical reaction of HER has followed two-electron ($2e^-$) transfer mechanism steps (i.e., Volmer and Heyrovsky). The electrochemical performance for the HER and electrochemical catalytic activity of the 2D monolayer Janus MoSSe have been described by calculating the change of Gibb's free energy ($\Delta G$) during HER in the Volmer reaction, $H^{\bullet-}$ migration step and $H_2$ formation energy in the Heyrovsky reaction step. The values of the $H^{\bullet-}$migration and Heyrovsky reaction barriers ($\Delta G$) are 3.93 kcal.mol$^{-1}$ and 5.61 kcal.mol$^{-1}$, respectively, (computed in gas phase), and the values of $\Delta G$ of both the reaction barriers are about 7.10 kcal.mol$^{-1}$ and 4.72 kcal.mol$^{-1}$, respectively, found in the solvent phase (i.e., water solution) calculations. Activation energy values are decreased due to the steadiness of rate determinant $H_2$ formation Heyrovsky step's transition state in the gas phase reaction and different electronegativity of the chalcogens (S and Se) in the asymmetric JTMD MoSSe. These lower values of activation energies, Tafel slope (59.16 mV/dec) and larger value of TOF facilitated the increase of the adsorption of hydrogen on the MoSSe catalytic surface. The transition state TS1 is alleviated by the better overlap (compared to the pristine 2D TMDs, such as $MoS_2$, $WS_2$, etc.) of atomic *s*-orbital of the $H_2$ molecules and the *d*-orbitals of the Mo atoms during the HER process. The overlap of the *s*-orbitals of the hydrogen atom attached with the Mo in the MoSSe JTMD and the water cluster ($3H_2O + H_3O^+$) seemed in the HOMO-LUMO Heyrovsky's transition state TS2 during $H_2$ formation has been found in the present study, and this better overlap of the atomic orbitals during the $H_2$ creation in the Heyrovsky's TS2 reduces the reaction barrier. This stabilization of the reaction limiting step in both the gas and solvent phases is a key for reducing both the $H^{\bullet-}$migration and Heyrovsky's reaction energy barriers, which results in a better electrocatalytic performance for HER compared to the ordinary TMDs. This mechanism offers a key intuition on why the 2D monolayer MoSSe JTMD material is an effective electrocatalyst for HER, i.e., the tuning of the *d*-orbital of the JTMDs overlaps with the *s*-orbital $H_2$ is of the most importance during the reactions and formation of the transition state. We can say that the Janus 2D monolayer MoSSe material has the potential for applications in renewable energy technology and hydrogen fuel cell that can be stored, transported, and used in a zero-emission fuel cell of combustion engine in near future. In conclusion, 2D layer asymmetric structures have been proven to be excellent materials as catalysts for the evolution of hydrogen. To design a more reactive, efficient, and high performance electrocatalyst for HER from exposed surfaces and edges of the 2D monolayer Janus TMD material, the researchers must concentrate on reducing both the $H^-$ migration and Heyrovsky reaction barriers.

## Author Contributions:

Dr Srimanta Pakhira built the entire concept of the present research work, and he computationally investigated the equilibrium structures and electronic properties of the 2D monolayer Janus MoSSe material. Dr. Pakhira explored the whole reaction pathways; transitions states and reactions barriers and he explained the HER mechanism by the DFT calculations. Quantum calculations and theoretical models were designed and performed by Dr Pakhira. Dr Pakhira wrote the whole manuscript and prepared all the tables and figures in the manuscript. Mr. Shrish Nath Upadhyay helped Dr Pakhira to organize the manuscript.

## Conflicts of Interest:

The authors have no additional conflicts of interest.

## AUTHOR INFORMATION

## Corresponding Author: Dr Srimanta Pakhira

Email: spakhira@iiti.ac.in or spakhirafsu@gmail.com

ORCID: 0000-0002-2488-300X**Corresponding Author**

**Dr. Srimanta Pakhira** − *Department of Physics, Indian Institute of Technology Indore (IIT Indore), Simrol, Khandwa Road, Indore, Madhya Pradesh 453552, India;*
*Department of Metallurgy Engineering and Materials Science, Indian Institute of Technology Indore (IITI), Simrol, Khandwa Road, Indore, Madhya Pradesh 453552, India;*
*Centre for Advanced Electronics (CAE), Indian Institute of Technology Indore, Simrol, Khandwa Road, Indore, Madhya Pradesh 453552, India;*
ORCID: orcid.org/0000-0002-2488-300X;
Email: spakhira@iiti.ac.in or spakhirafsu@gmail.com

**Author**

**Mr. Shrish Nath Upadhyay** − *Department of Metallurgy Engineering and Materials Science (MEMS), Indian Institute of Technology Indore (IIT Indore), Simrol, Khandwa Road, Indore, Madhya Pradesh 453552, India*;
ORCID: orcid.org/0000-0003-0029-4160.## Acknowledgment:

We acknowledge the funding and technical support from the Science and Engineering Research Board-Department of Science and Technology (SERB-DST), Govt. of India under the Grant No. ECR/2018/000255. This research work is fiscally supported by the SERB-DST, Govt. of India under the Grant No. ECR/2018/000255. Dr. Srimanta Pakhira is grateful for the financial support from the SERB-DST, Govt. of India under the scheme number ECR/2018/000255. Dr. Pakhira thanks to the SERB for providing highly prestigious Ramanujan Faculty Fellowship under the scheme number SB/S2/RJN-067/2017 and Core Research Grant (CRG), SERB-DST, Govt. of India under the scheme number CRG/2021/000572. Mr. Upadhyay thanks Indian Institute of Technology Indore, MHRD, Govt. of India for providing his doctoral fellowship. We acknowledge to Ms. Stephanie F. Marxsen, Department of Chemical and Biomedical Engineering, College of Engineering (COE), Florida State University (FSU), Tallahassee, FL, United States America (USA) for valuable discussions. We would like to acknowledge the Editor for offering the opportunity to revise our work after addressing the reviewers' concerns. We thank the respected reviewers for their valuable comments and suggestions.40

## TOC: Graphical Abstract

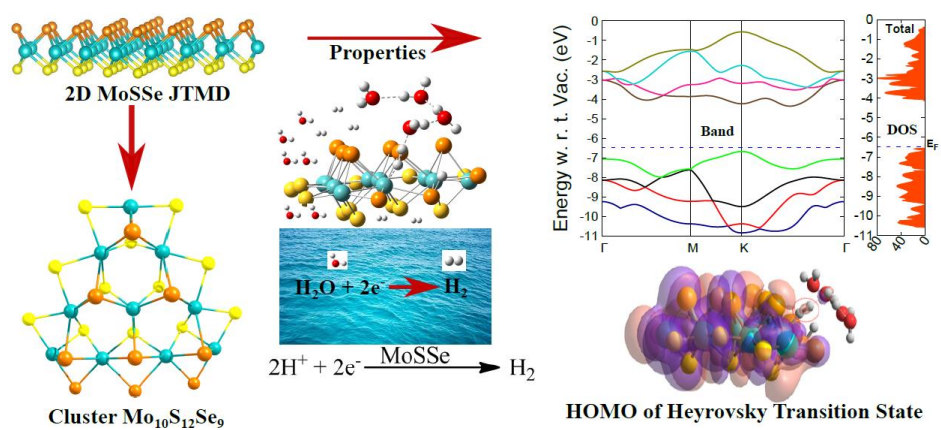